\documentclass[twocolumn,prb,aps,showpacs,superscriptaddress]{revtex4}
\usepackage{graphics}
\usepackage{epsfig}
\usepackage{latexsym}
\newcommand{\av}[1]     {\langle #1 \rangle}
\newcommand{\Av}[1]     {\left\langle #1 \right\rangle}

\newcommand{\eqn}[1]    {(\ref{#1})}
\newcommand{\fig}[1]	{Fig. \ref{#1}}
\newcommand{\subbox}[1]	{{\mbox{\scriptsize #1}}}

\def\etal	{{\em et al.~}}

\def\bi         {\begin{itemize}}
\def\ei         {\end{itemize}}
\def\benu	{\begin{enumerate}}
\def\eenu	{\end{enumerate}}
\def\bmat       {\left[ \begin{array}}
\def\emat       {\end{array} \right]}
\def\beq	{\begin{equation}}
\def\eeq	{\end{equation}}
\def\beqn       {\begin{eqnarray*}}
\def\eeqn       {\end{eqnarray*}}
\def\beqa       {\begin{eqnarray}}
\def\eeqa       {\end{eqnarray}}
\def\bquote	{\begin{quote}}
\def\equote	{\end{quote}}
\def\f          {\frac}
\def\bwide	{\begin{widetext}}
\def\ewide	{\end{widetext}}

\def\a          {\alpha}

\def\d          {\delta}
\def\e          {\epsilon}

\def\g          {\gamma}
\def\k          {\kappa}
\def\l		{\lambda}
\def\m          {\mu}

\def\s          {\sigma}
\def\t          {\tau}
\def\th         {\theta}
\def\ve		{\varepsilon}

\def\w          {\omega}
\def\x          {\xi}

\def\D          {\Delta}

\def\Si          {\Sigma}

\def\W          {\Omega}

\def\bdf        {{\bf f}}

\def\bp         {{\bf p}}

\def\br         {{\bf r}}

\def\bv         {{\bf v}}

\def\bF         {{\bf F}}
\def\bG         {{\bf G}}
\def\bH         {{\bf H}}

\def\bel        {{\mbox{\boldmath$\ell$}}}

\def\bgam	{{\mbox{\boldmath$\g$}}}
\def\brho       {{\mbox{\boldmath$\rho$}}}

\def\bDel       {{\mbox{\boldmath$\Delta$}}}

\def\cA		{{\mathcal{A}}}
\def\cB         {{\mathcal{B}}}

\def\cF		{{\mathcal{F}}}
\def\cG		{{\mathcal{G}}}

\def\cO		{{\mathcal{O}}}
\def\cP		{{\mathcal{P}}}
\def\cR		{{\mathcal{R}}}

\def\bcF	{{\mbox{\boldmath$\cF$}}}

\def\rarr	{\rightarrow}

\def\uparr	{\uparrow}
\def\dnarr	{\downarrow}
\def\dag	{\dagger}

\def\im		{{\mbox{Im}}}
\def\re		{{\mbox{Re}}}

\def\sgn	{{\mbox{sgn}~}}

\begin{document}
\title{Odd triplet pairing effects induced by interface spin-flip scatterings:
critical temperature of superconductor/ferromagnet bilayers}

\author{Hyeonjin Doh}
\email{clotho@phys1.skku.ac.kr} \affiliation{Department of
Physics, BK21 Physics Research Division, Institute for Basic
Science Research, Sung Kyun Kwan University, Suwon 440-746,
Korea.}
\author{Han-Yong Choi}
\email{hychoi@skku.ac.kr} \affiliation{Department of Physics,
BK21 Physics Research Division, Institute for Basic Science
Research, Sung Kyun Kwan University, Suwon 440-746, Korea.}
\affiliation{Asia Pacific Center for Theoretical Physics, Pohang
790-784, Korea.}
\begin{abstract}
The superconducting critical temperature $T_C$ of a
superconductor/ferromagnet (S/F) bilayer with spin-flip
scatterings at the interface is calculated as a function of the
ferromagnet thickness $d_F$ in the dirty limit employing the
Usadel equation. The appropriate boundary conditions from the
spin-flip scatterings at the S/F interface are derived for the
Usadel equation which includes the spin triplet pairing components
as well as the spin singlet one. The spin-flip processes induce
the spin triplet pairing components with $s$-wave in momentum and
odd symmetry in frequency from the $s$-wave singlet order
parameter $\Delta$ of the superconductor region. The induced
triplet components alter the singlet order parameter in the
superconductor through boundary conditions at the interface and,
consequently, change the $T_C$ of an S/F bilayer system. The
calculated $T_C(d_F)$, like the case of no spin-flips, shows
non-monotonic behavior which typically decreases as $d_F$ is
increased from 0 and shows a shallow minimum and then saturates
slowly as $d_F$ is further increased. It is well established that
as the interface resistance (parameterized in terms of $\gamma_b$)
is increased, the $T_C$ is increased for a given $d_F$ and the
non-monotonic feature in $T_C(d_F)$ is strongly suppressed. As
the spin flip scattering (parameterized in terms of $\gamma_m$) is
increased, on the other hand, the $T_C$ is also increased for a
given $d_F$, but the non-monotonic feature in $T_C(d_F)$ is less
suppressed or even enhanced, through the formation of the spin
triplet components.
\end{abstract}
\maketitle
\section{Introduction}
\label{sec:intro} The superconductivity and ferromagnetism are
two competing orders: The former prefers a spin anti-parallel
state and the latter a spin parallel state. They can coexist and
exhibit interesting interplay effects only in a very narrow range
of parameters. The effects of their competition and interplay,
therefore, can be more conveniently studied when the interactions
responsible for the two orders are confined to spatially separate
regions of a system like a superconductor/ferromagnet (S/F)
junction. When a superconductor is brought into a contact with a
ferromagnet, the two competing orders influence each other in the
vicinity of the interface on a spatial scale of the order of the
coherence lengths. These phenomena of mutual influences are
referred to as ``proximity effects''. The proximity effects of a
superconductor/normal metal (S/N) junction show up as a
penetration of the superconducting pairing amplitude into the
normal metal region with an exponential decay
\cite{Werthamer63pr,deGennes64rmp}. In an S/F junction, on the
other hand, the superconducting pairing amplitude $\Psi(x)$ in the
ferromagnet region does not simply decay exponentially but also
oscillates. This oscillation shows up because the electrons
forming Cooper pairs feel a distinct potential depending on their
spin due to the exchange field in the ferromagnet, which leads to
a finite momentum of a Cooper pair similar to the
``Fulde-Ferrel-Larkin-Ovchinikov state'' (FFLO state) in bulk
materials \cite{Fulde64pr,Larkin65jetp}.

The same physics behind this pairing amplitude oscillation
manifests itself in a number of different contexts: For instance,
the non-monotonic dependence of $T_C$ of S/F systems on the
ferromagnet thickness $d_F$ and the ``$\pi$-state'' of  S/F/S
junctions, among others. Here, the $T_C$ is the superconducting
critical temperature of an S/F junction where the current is
parallel to the interface. Then, the $T_C$ corresponds to the
highest temperature at which the superconducting order parameter
$\Delta(x)$ becomes non-vanishing at least at one point within the
junction. The $\Delta(x)$ defined in Eq.\ (\ref{phix}) and
$\Psi(x)$ are related as
 \beqa
\Delta(x) = \lambda \Psi(x),
 \eeqa
where $\lambda$ is the superconducting pairing interaction. The
$\pi$-state refers to a case where two superconductors of a S/F/S
junction separated by a ferromagnet of an appropriate thickness
have the phase difference of $\pi$ in a ground state without any
external gauge field. That is, the order parameters of two
superconducting regions have the opposite signs. The magnetic
coherence length, which determines the oscillation length and the
penetration depth of the pairing amplitude $\Psi(x)$ in the
ferromagnet, the ferromagnet thickness $d_F$ of minimum $T_C$ in
$T_C(d_F)$, the ferromagnet thickness of the $\pi$-junction, and
so on, is given roughly by
 \beqa
 \label{xiF0}
\xi_F^\subbox{ex} = \left\{
\begin{array}{ll}
\f{\hbar v_F}{\pi E_\subbox{ex}} & {\rm for ~clean~ limit,} \\
 \sqrt{\f{\hbar v_F \ell}{\pi E_\subbox{ex}}} & {\rm for~ dirty~ limit,}
\end{array} \right.
 \eeqa
where $E_\subbox{ex}$ is the exchange energy and $\ell$ is the
mean free path. For a ferromagnet, $E_\subbox{ex}$ is about
thousands Kelvin, which gives a $\xi_F^\subbox{ex}$ of a few
nanometers. Studies of the oscillatory behavior of S/F junctions,
therefore, require a fabrication technique of a nanometer scale
control of the F thickness.

Due to the difficulties in the fabrication, serious studies of S/F
junctions were initiated by theoretical works in the late 70's.
Bulaevskii \etal considered magnetic impurities in Josephson
junction \cite{Bulaevskii77sjltp}. Buzdin \etal predicted the
critical current \cite{Buzdin82jetpl} and $T_C$ oscillation
\cite{Buzdin90jetpl} as a function of $d_F$, and studied the
effects of ferromagnet layer in Josephson junction
\cite{Buzdin91jetpl}. Following these theoretical works, many
experimental efforts have been made to test the pairing amplitude
oscillation in the ferromagnet. One of them is to measure the
$T_C$ vs.\ $d_F$ of S/F bilayers. The measured $T_C$ showed a
non-monotonic dependence on $d_F$ \cite{Jiang95prl}, although
there were some contradictory reports \cite{Muhge96prl}. Another
line of research is to verify the $\pi$-state. Quite a few
experiments reported the $\pi$-state: tunneling spectroscopy
\cite{Kontos01prl}, temperature dependence \cite{Ryazanov01prl},
and thickness dependence of the critical current
\cite{Kontos02prl}, and even phase sensitive measurements
\cite{Guichard03prl}. Through all the above experiments, the
pairing amplitude oscillation in the superconductor-ferromagnet
hetero structures is now well established.

The advancement of the experimental technique of the S/F systems,
on the other hand, demanded more detailed theoretical analysis of
the systems to include those effects hitherto not considered such
as the inhomogeneity of the magnetization. For example, there
were some measurements that reported much longer penetration
lengths in S/F junctions \cite{Giroud98prb,Petrashov99prl}. The
$T_C$ difference between the parallel and anti-parallel
magnetizations of the two ferromagnets sandwiching the
superconductor in F/S/F junctions is smaller by the factor of
$10^2$ than the theory predicts \cite{Gu02prl}. For filling the
gap between the theories and experiments, the spin-orbit
scattering in the ferromagnet \cite{Demler97prb}, or the triplet
components \cite{Bergeret01prl,Edelstein03jetpl,Eschrig03prl,
Volkov03prl,Fominov03jetpl,Bergeret03prb} began to be considered
by several groups. It should be pointed out that there is $no$
pairing interaction in the triplet channel in the S or F region
because conventional $s$-wave superconductors were considered for
S/F bilayer junctions.

\begin{figure}[htb]
\epsfxsize=7cm \epsffile{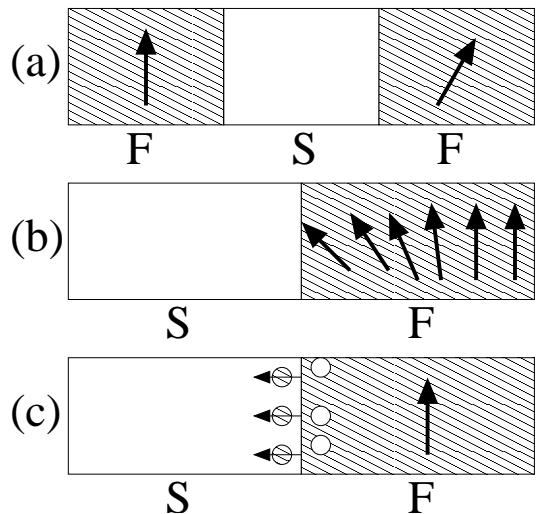} \caption{Some schematic
pictures of cases which can induce the triplet pairing
components. (a) non-collinear magnetizations of two ferromagnets
separated by a superconducting layer, (b) rotation of
magnetization in a ferromagnet layer, and (c) spin-flip scattering
at the interface of the magnetic and the non-magnetic layers due
to mixing of two materials or spin-orbit scattering.
\label{fig:triplet_sf}}
\end{figure}

The triplet pairing components are induced in a hybrid junction
of S and F when a single quantization axis for spin can $not$ be
defined for the whole junction. Such situations arise and the
triplet pairings were studied for (a) the non-collinear alignment
of the magnetic moments of two F's in a F/S/F structure
\cite{Volkov03prl,Fominov03jetpl,Bergeret03prb}
[Fig.~\ref{fig:triplet_sf} (a)], (b) a rotation of the
magnetization direction within a finite thickness in a F region
of S/F bilayers \cite{Bergeret01prl} [Fig.~\ref{fig:triplet_sf}
(b)], (c) spin-orbit scattering near the interface between
non-magnetic and magnetic materials in S/F structures
\cite{Edelstein03jetpl}, and (d) spin mixing near the interfaces
of S/F/S structure where F is a half-metal \cite{Eschrig03prl}.
The works of (a)-(b) considered the dirty limit cases. In the
dirty limit, the non-$s$ wave components are strongly suppressed,
and the so called ``odd'' frequency triplet superconductivity is
more robust than the $p$-wave triplet superconductivity. The odd
triplet pairing has even symmetry in spin and momentum spaces but
odd symmetry in frequency, and was first proposed by Berezinskii
\cite{Berezinskii74jetpl} in 1974 in the context of $^3$He
superfluidity. The odd frequency triplet pairings, interestingly,
can be realized in hybrid junctions of S and F as discussed in
(a)-(b) which employed the Usadel equation in the dirty limit
\cite{Usadel70prl}. The work (c)-(d) considered the $p$-wave pairing
in the clean limit.

In this paper, we considered an S/F bilayer as in (b) and (c),
and studied the effects of the induced triplet pairing components
in S/F bilayers by employing the Usadel equation. The difference
is that we consider the spin flip scatterings at the interface.
Technically, this calls for new boundary conditions at the
interface of an S/F junction. We derived such boundary conditions
accommodating the spin-flip scatterings at the interface. The
work of (b) simplified and bypassed the problem of the new
boundary conditions by considering an artificial rotation of the
magnetization of a fixed amplitude in the F region. On the other
hand, we consider an intrinsic magnetic inhomogeneity at the
interface of S/F junctions as depicted in
Fig.~\ref{fig:triplet_sf} (c). The Usadel equation is appropriate
in the dirty limit and is widely employed for analyzing various
S/N, S/F, and S/N/F systems since various junctions of S, N and F
films belong to the dirty limit. Here, we calculate the critical
temperature $T_C$ as a function of the ferromagnet thickness
$d_F$ of S/F bilayers including the triplet pairing components in
addition to the singlet one.



This paper is organized as follows. In Sec.\ \ref{sec:model}, we
will introduce the notations and write down the Usadel equation
and boundary conditions that need to be solved. For the boundary
conditions of Usadel equation with spin-flip scatterings, the
boundary conditions of the Eilenberger equation is first derived
in the Appendix \ref{sec:BC_Eilenberger} and, in Appendix
\ref{sec:BC_Usadel} and \ref{sec:normal_green}, the detailed
derivation of the boundary conditions for Usadel equation is
collected. This will render the paper more easily accessible to
the less technically inclined. Then, the Usadel equation with the
appropriate boundary conditions is mapped onto an eigenvalue
problem [See Eq.\ (\ref{eqn:discrete}) below.] whose detailed
derivation is given in Appendix \ref{sec:TcCalc}. Actual
calculations of $T_C$ constitute setting up the matrix $K_{ij}$
of Eq.\ (\ref{eqn:discrete}) and finding its smallest eigenvalue
self-consistently. The self-consistency is achieved via
iterations. In Sec.\ \ref{sec:result}, the results of numerical
$T_C$ calculations of S/F bilayers with the spin-flip scattering
induced triplet components will be presented. Overall, the $T_C$
of S/F bilayers with the spin-flip scatterings, like the more
familiar case of no spin-flips, typically decreases initially as
$d_F$ is increased from 0, then shows a minimum at a finite $d_F$,
then increases slowly and saturates to
$T_C^\ast=T_C(d_F\rarr\infty)$ as $d_F$ is further increased. As
the spin flip scatterings are increased, compared with the
corresponding no spin flip case, (a) the $T_C$ is increased with
decreasing non-monotonic behavior, but (b) the relative
non-monotonic feature in $T_C(d_F)$ with respect to $T_C^\ast$ is
enhanced (the increase of $T_C^\ast$ is faster than the minimum
$T_C$), through the formation of spin triplet components.
Finally, we have summary and some concluding remarks in Sec.\
\ref{sec:conclusion}.
\section{Formulation of S/F bilayer}
\label{sec:model}
In this section, we will present the Usadel
equation with the induced odd-frequency $s$-wave triplet pairing
components in addition to the dominant singlet pairing order
parameter, boundary condition at the S/F interface, and
calculation of $T_C$ by mapping onto an eigenvalue problem. The
detailed derivations of the boundary conditions and mapping of
$T_C$ to the eigensystem are collected in the Appendix
\ref{sec:BC_Usadel}, and \ref{sec:TcCalc}, respectively. We
consider the dirty limit case which almost all experiments belong
to.

The Usadel equation which considers both the singlet and triplet
pairing components can be written using the anomalous function
in spin space as follows:
\beqa
\label{Usadel:s&t}
\xi^2\pi
T_C\f{\partial^2}{\partial x^2}\hat{F}(x,i\w)= |\w|\hat{F}(x,i\w)
-\hat{\D}(x)\\
-i\ \sgn(\w)\left(
\hat{H}_\subbox{ex}\hat{F}(x,i\w)
-\hat{F}(x,i\w) \hat{H}_\subbox{ex}^\ast
\right),
\nonumber
\eeqa
where $\hat{F}$ is the Usadel function from anomalous Green's function
of $2\times 2$ matrix in spin space and $\hat{\D}$ is the singlet pairing
order parameter.
\beq
\hat{F}=\bmat{cc}
F_{\uparr\uparr}&F_{\uparr\dnarr}\\
F_{\dnarr\uparr}&F_{\dnarr\dnarr}
\emat,~~~
\hat{\D} = \bmat{cc}
0 & \D \\
-\D & 0 \emat \eeq $\hat{H}_\subbox{ex}$ is also $2\times 2$
matrix which denotes the exchange field in the ferromagnet. \beq
\hat{H} = H_\subbox{ex}^x \s_x + H_\subbox{ex}^y \s_y +
H_\subbox{ex}^z \s_z \eeq where $\s_i$'s are the Pauli matrices.
To show the coupling between the singlet and triplet pairing state
clearly, the following vector form of the Usadel equation can be
convenient in numerical calculations.
 \beqa
 \label{Usadel:vector}
\xi^2\pi
T_C\f{\partial^2}{\partial x^2}\bF(x,i\w)&=& |\w|\bF(x,i\w)
-\bDel(x) \\
&&-i\ \sgn(\w)\ \bH_\subbox{ex}\cdot \bF(x,i\w), \nonumber
 \eeqa
where $\w=2\pi T(n+1/2)$ is the Matsubara frequency, $D=v_F^2
\tau/3$ is the diffusion constant, and
 \beqa
 \label{xiTc}
\xi = \sqrt{D/2\pi T_C}
 \eeqa
is the coherence length in dirty limit. The $\bF$ and $\bDel$ are four component
vectors and $\bH_\subbox{ex}$ is a $4\times 4$ exchange field
tensor given, respectively, by
\beqa
&&\bF(x,i\w) =
\bmat{c}F_s(x,i\w)\\F_{tx}(x,i\w)\\F_{tz}(x,i\w)\\F_{t0}(x,i\w)\emat,~~
\bDel(x) = \bmat{c}\D(x)\\0\\0\\0\emat,\nonumber\\
&&\bH_\subbox{ex} =
\bmat{cccc}
0 & -H_\subbox{ex}^z & - H_\subbox{ex}^x & -i H_\subbox{ex}^y\\
-H_\subbox{ex}^z & 0 & 0 & 0 \\
-H_\subbox{ex}^x & 0 & 0 & 0 \\
i H_\subbox{ex}^y&0 &0 & 0\emat. \label{eqn:def-bfbDel} \eeqa
$\D(x)$ is the singlet $s$-wave pairing order parameter and a
function of $x$ which represents the coordinate perpendicular to
the interface between S and F. The $F_\a$'s are transformations
of the $F_{\sigma\sigma'}$ by expanding $\hat{F}$ on the basis of
the Pauli matrices as shown in Eq.\ \eqn{eqn:pauli-expand} in
Appendix \ref{sec:BC_Usadel}. The transformations are given as:
 \beqa
F_{s}= \f{1}{2} (F_{\uparr\dnarr}-F_{\dnarr\uparr}),~~ F_{tx} =
\f{1}{2}(F_{\uparr\dnarr}+F_{\dnarr\uparr}),\nonumber \\ F_{tz} =
\f{1}{2}(F_{\uparr\uparr}-F_{\dnarr\dnarr}),~~ F_{t0}=
\f{1}{2}(F_{\uparr\uparr}+F_{\dnarr\dnarr}).
 \eeqa
The $F_s$ is the singlet pairing component and the others are the
triplet pairing components. The self-consistency relation is
given by
 \beq \label{phix} \Delta(x) =\lambda\pi T_C g(\epsilon_F)
\sum_{\omega_n} F_s (x,i\omega_n),
 \eeq
where $\lambda$ is the superconducting pairing interaction and
$g(\epsilon_F)$ is the density of states per spin at the Fermi
level. Since we assume that $\l$ is zero at ferromagnet, we
should note that the superconducting order $\D$ is also zero at
ferromagnet and that only the pairing amplitude $\Phi$ can
penetrate into.

\begin{figure}[hbt]
\epsfxsize=8cm
\epsffile{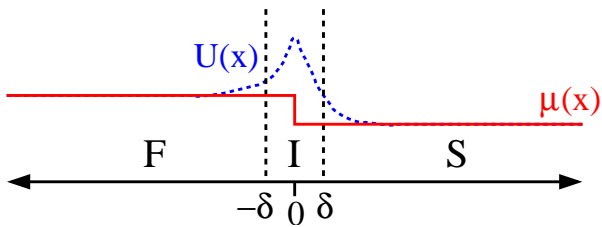}
\caption{
The hypothetical
layer (I) with an infinitesimal thickness $2\d$ between
superconductor (S) and ferromagnet (F) which are described as
a spin-dependent potential $U(x)$ which will give a interface resistance and
a spin-flip scattering.
$\m(x)$ stands for a local chemical potential.
\label{fig:interface}}
\end{figure}

The Usadel equation of Eq.\ \eqn{Usadel:s&t} or
\eqn{Usadel:vector} must be supplemented by a set of appropriate
boundary conditions. The system we consider is an S/F junction
where the interface between S and F has the potential (without
spin-flip) and spin-flip scatterings due to magnetic inhomogeneity
as shown in Fig.\ \ref{fig:interface}. Therefore, the interface
potential is represented by $\d$ functions with the spin
dependency as
 \beq
\label{eqn:interface_delta} U(x) = \left(V_0\s_0 +
V_x\s_x+V_y\s_y+V_z\s_z\right)\d(x).
 \eeq
For simplicity, we can set $V_0$ and $V_x$ to be the only
non-zero parameters among the spin dependent $\d$-function, when
the magnetization of ferromagnetic layer is in $z$-direction. The
spin-flip scatterings were modeled in terms of the $\d$-function-
like magnetization at the interface. In real situations, this
will be local inhomogeneities of the magnetization within a very
short range near the interface. Within the Usadel formulation, the
boundary conditions using the similar notation with Eq.\
(\ref{Usadel:vector}) can be written as
 \beqa
\label{eqn:bc-spinflip1} \bF^S-\bF^F = \left[\hat{\g}_b-i\ {\rm
sgn}(\w) \hat{\g}_m\right]
\xi_F\f{\partial}{\partial x}\bF^F , \\
\xi_S\f{\partial}{\partial x}\bF^S
-\g\xi_F\f{\partial}{\partial x}\bF^F =0,
\label{eqn:bc-spinflip2}
 \eeqa
where the dimensionless parameters are defined by
 \beqa
\label{eqn:bc-def} \g = \f{\rho_S\xi_S}{\rho_F\xi_F},~
 \hat{\g}_b &=& \bmat{cccc}
\g_b + \g_1 & 0 & 0 & 0 \\
0 & \g_b - \g_1 & 0 & \g_2 \\
0 & 0 & \g_b + \g_1 & 0 \\
0 & \g_2 & 0 & \g_b - \g_1 \emat,\nonumber\\
 \hat{\g}_m &=& \bmat{cccc}
0 & 0 & \g_m & 0 \\
0 & 0 & 0 & 0 \\
\g_m & 0 & 0 & 0 \\
0 & 0 & 0 & 0
\emat.
 \eeqa
The $\gamma$ is the ratio of S and F bulk properties, and
$\g_b$ represents the contact quality which is expressed as
$\f{\cR\cA}{\rho_F\xi_F}$ without spin-flip scatterings. $\cal A$
and $\cal R$ are the interface area and resistance, respectively.
%
$\gamma_m$ is the spin-flip strengths between S and F which is
proportional to $V_x$ in \eqn{eqn:interface_delta}. $\g_1$ and
$\g_2$ comes from the higher order term of $V_x$ and $V_0$ which
can be ignored when the scattering strength is weak. See Appendix
\ref{sec:BC_Usadel} for details.

The Usadel equation of Eq.\ (\ref{Usadel:vector}) with the
boundary conditions of Eqs.\ (\ref{eqn:bc-spinflip1}) and
(\ref{eqn:bc-spinflip2}) was mapped onto an eigenvalue problem and
solved numerically to determine the $T_C$. The $T_C$ defined here
can be determined experimentally by measuring the resistance with
the current parallel to the interface. For simplicity, we take the
magnetization at the interface along the $x$-axis and that of
ferromagnet along the $z$-axis.

We follow Fominov \etal\cite{Fominov01jetpl,Fominov02prb} to get
the numerical solution of the Usadel equation. We first separate
the anomalous Green's function $F_\a(x,i\w)$ into the even and odd
symmetry parts in the frequency space
 \beq
F_\a^{(\pm)}(x,i\w) = F_\a(x,i\w)\pm F_\a(x,-i\w),
 \eeq
where $\a=s,~tx,~tz,~\mbox{and }t0$.
Since all the anomalous Usadel functions are $s$-wave pairing,
we choose the even symmetry in frequency for the singlet component, and
the odd symmetry for the triplet components.
Thus we define a 4-component vector $\bcF$ as
 \beq
\label{eqn:def-bF}
\bcF(x,i\w)\equiv
\bmat{c}
F_{s}^{(+)}(x,i\w)\\
F_{tx}^{(-)}(x,i\w)\\
F_{tz}^{(-)}(x,i\w)\\
F_{t0}^{(-)}(x,i\w)
\emat.
 \eeq
Here, the triplet components, $F_{tx}$, $F_{tz}$ and $F_{t0}$,
are odd in the frequency and even in the momentum space as we
mentioned in Sec.\ \ref{sec:intro}. This special type of triplet
pairing is the odd triplet pairing proposed by Berezinskii
\cite{Berezinskii74jetpl}. It will be much more robust than a
$p$-wave triplet pairing to disorders due to the momentum
independent gap feature.

\begin{figure}[hbt]
\epsfxsize=8cm \epsffile{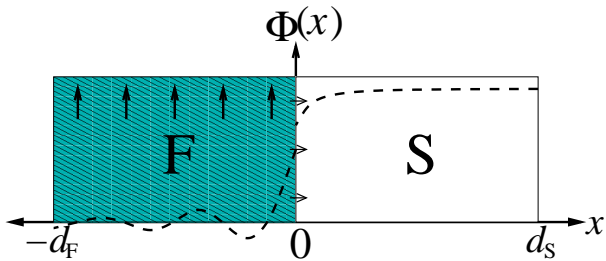} \caption{A schematic
plot of the pairing amplitude $\Phi(x)$ of Eq.\ (\ref{phix})
with the dotted line for an S/F bilayer. The magnetization in the
ferromagnet region is taken in the $z$-direction and the
magnetization at the S and F interface is in the $x$-direction.
\label{fig:sflayer}}
\end{figure}

As shown in Fig.~\ref{fig:sflayer}, for an S/F bilayer, we took
the superconductor region in $0<x<d_S$, the ferromagnet in
$-d_F<x<0$, and the spin-flip scatterings at the interface of
$x=0$. In the S region, there is no exchange field and only
pairing order parameter $\D(x)$ exists. Therefore, the Usadel
equation of \eqn{Usadel:vector} can be written for the S region as
 \beq \pi T_C\xi_S^2\f{\partial^2}{\partial x^2}
\bcF^{S}(x,i\w) = |\w| \bcF^{S}(x,i\w) - 2\bDel(x),
\label{eqn:bi-usadel_s}
 \eeq
for $0<x<d_S$.
For the F region, on the other hand, there is no pairing
interaction and only the exchange field term appears. The Usadel
equation is then written as
 \beq
\pi T_C\xi_F^2\f{\partial^2}{\partial x^2} \bcF^{F}(x,i\w)
= (|\w| -i \hat{\bH}_\subbox{ex} )\bcF^{F}(x,i\w),
\label{eqn:bi-usadel_f}
 \eeq
for $-d_F <x<0$. The coherence length $\xi_F$ in the F region is
given by Eq.\ (\ref{xiTc}) as
 \beqa
\xi_F = \sqrt{\frac{D_F}{2\pi T_C}},
 \eeqa
where $D_F$ is the diffusion constant of F and $T_C$ is the
superconducting transition temperature of the S/F junction. The
$\xi_F$ should be distinguished from the $\xi_F^\subbox{ex}$ of
Eq.\ (\ref{xiF0}).
$\xi_F^\subbox{ex}$ corresponds to the actual penetration depth of
singlet pairing amplitude in F,
but $\xi_F$ is the pure penetration depth
without considering the exchange field as in normal metal.
Later, this $\xi_F$ will turn out to be a penetration depth of
triplet components, $F_{tz}$ and $F_{t0}$,
which is unaffected by the exchange energy.
The boundary conditions of Eqs.\
(\ref{eqn:bc-spinflip1}) and (\ref{eqn:bc-spinflip2}) can be
written in terms of $\bcF(x,i\w)$ of Eq.\ (\ref{eqn:def-bF}) as
follows:
 \beqa
&&\f{\partial}{\partial x}\bcF^S(d_S)=0,~~~
\f{\partial}{\partial x}\bcF^F(-d_F)=0,
\label{eqn:bi-bc1}\\
&&\bcF^{S}(0)-\bcF^{F}(0)= \hat{\bgam}_b
\xi_F\f{\partial}{\partial x}\bcF^{F}(0),
\label{eqn:bi-bc2}\\
&&\xi_S\f{\partial}{\partial x}\bcF^{S}(0)
-\g\xi_F\f{\partial}{\partial x}\bcF^{F}(0) =0,
\label{eqn:bi-bc3}
 \eeqa
where
\beq
\label{eqn:gamma_B}
\hat{\bgam}_b = \bmat{cccc}
\g_b & 0 & -i\g_m & 0 \\
0 & \g_b & 0 & 0 \\
-i\g_m & 0 & \g_b & 0 \\
0 & 0 & 0 & \g_b \emat. \eeq Here, $\g_1$ and $\g_2$ of Eq.\
(\ref{eqn:bc-def}) are ignored for the weak scattering rate at the
interface.

As detailed in Appendix \ref{sec:TcCalc}, Eqs.\ (\ref{eqn:bi-usadel_s}),
(\ref{eqn:bi-usadel_f}), (\ref{eqn:bi-bc1}), (\ref{eqn:bi-bc2}),
and (\ref{eqn:bi-bc3}) can be solved for the $T_C$ of a S/F
bilayer, and the $T_C$ equation with respect to the bulk
transition temperature $T_{C0}$ may be cast into the form
 \beq \label{eqn:discrete}
\D_i\ln\left(\f{T_{C0}}{T_C}\right)=\sum_{j}^{N} K_{ij} \D_j ,
 \eeq
where the order parameter $\Delta$ and kernel $K$ are defined on a
discrete grid as
 \beqa \label{eqn:kernel}
\D_i \equiv \D\left(\f{d_S}{N}i\right), \nonumber \\ K_{ij}\equiv
2\pi T_C\sum_{\w_n>0}
\left[\f{\d_{ij}}{|\w_n|}-G_{ij}(i\w_n)\right], \nonumber \\
G_{ij}(i\w_n)\equiv\f{d_S}{N}G\left(\f{d_S}{N}i,\f{d_S}{N}j,i\w_n\right).
 \eeqa
Here, $N$ is the number of the discrete grids in the integral
domain between 0 and $d_S$. The $T_C$ is determined by the
smallest eigenvalue of the matrix $K$ given by \eqn{eqn:kernel},
which gives the largest $T_C$. However, the kernel $K_{ij}$ is
$T_C$ dependent through the Matsubara frequencies $\w_n = 2\pi T_C
(n+1/2)$. Therefore, we solve the Eq.\ (\ref{eqn:discrete})
iteratively until it gives a self-consistent solution for $T_C$.
\section{Numerical Results}
\label{sec:result}
The interface parameter of the boundary
condition, $\hat{\bgam}_b$ in \eqn{eqn:bi-bc2}, contains diagonal
and off-diagonal elements as given in \eqn{eqn:gamma_B}. The
diagonal element, $\g_b$, corresponds to the interface resistance
and the off-diagonal element, $\g_m$, to the scattering between
singlet and triplet pairing components.
The relation between the parameters $\gamma_b$, $\gamma_m$ and
the more microscopic processes like the potentials $V_0$, $V_x$ is
complicated and depends on detailed mechanism. In the present
work, therefore, we take the $\g_b$ and $\g_m$ in
\eqn{eqn:gamma_B} as independent parameters characterizing the
S/F interface without explicit references to more microscopic
processes. For the present microscopic consideration in terms of
$V_0$ and $V_x$, the connection between the interface parameters
and the microscopic potentials can be found in Appendix B.

The Usadel equations in Eqs. \eqn{eqn:bi-usadel_s} and 
\eqn{eqn:bi-usadel_f} with the boundary condition in Eqs. 
\eqn{eqn:bi-bc1}, \eqn{eqn:bi-bc2}, and \eqn{eqn:bi-bc3}, can be 
expressed by the seven dimensionless parameters, $d_S/\xi_S$, 
$d_F/\xi_F$, $T_C/T_{C0}$, $H_\subbox{ex}/T_{C0}$, $\g$, $\g_b$, 
and $\g_m$. All the calculations for $T_C/T_{C0}$ are done and 
shown in the figures with respect to the other six independent 
parameters. The numerical results were calculated based on the 
parameters in Ref.\ \onlinecite{Fominov02prb}, which are $T_{C0} 
= 7.0$ K, $\xi_S = 8.9$ nm, $\xi_F = 7.6$ nm, $\rho_S = 7.5$ 
$\m\W$cm, $\rho_F = 60.0$ $\m\W$cm, and $d_S = 11$ nm.

\begin{figure}[hbt]
\epsfxsize=8cm
\epsffile{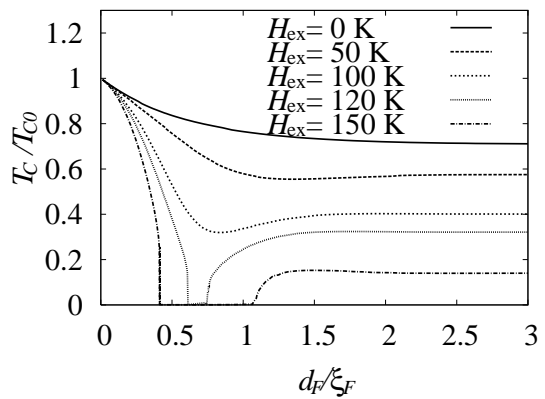}
\caption{$T_C/T_{C0}$
vs.\ $d_F/\x_F$ for the various $H_\subbox{ex}$ without interface
resistance and spin-flip
scattering at the interface, i.e.\ $\g_b=0.0$ and $\g_m = 0.0$.
\label{fig:tcdfsf}}
\end{figure}
First, we show the typical behavior of $T_C(d_F)$ in S/F bilayer
in \fig{fig:tcdfsf} as well as $T_C(d_N)$ in S/N bilayer (solid
line) which is S/F bilayer with zero exchange field,
$H_\subbox{ex}=0.0$. For S/N bilayer, $T_C$ decreases
monotonically as the thickness of normal metal layer increases.
When the exchange field is turned on, the $T_C$'s, the dotted
curves in \fig{fig:tcdfsf}, show typical behaviors of $T_C(d_F)$
for S/F bilayers. As $d_F$ is increased from 0, $T_C$ decreases
fast initially and shows a minimum, and then increases slightly
and saturates as $d_F$ is further increased. This non-monotonic
dependence of $T_C$ on $d_F$ is due to the oscillation of the
pairing amplitude in ferromagnet, known as FFLO state. The
thickness of minimum $T_C$ in $T_C(d_F)$ is almost corresponding
to the dirty limit value of $\xi_F^\subbox{ex}$ in Eq.\
\eqn{xiF0}. For the same reason, the minimum $T_C$ can be also
denoted as the transition from $0$-state to $\pi$-state of
pairing amplitude in F. As shown in Eq.\ \eqn{xiF0}, the position
of minimum $T_C$ is shifted to left as $H_\subbox{ex}$ increases.
For a strong enough exchange energy, $T_C$ is strongly suppressed
and goes to zero as $d_F$ increases.\cite{Fominov02prb} For a
properly strong exchange field, moreover, the reentrant behavior
of superconductivity occurs as the cases of $H_\subbox{ex} = 120$
K and 150 K shown in \fig{fig:tcdfsf}.

\begin{figure}[hbt]
\epsfxsize=8cm \epsffile{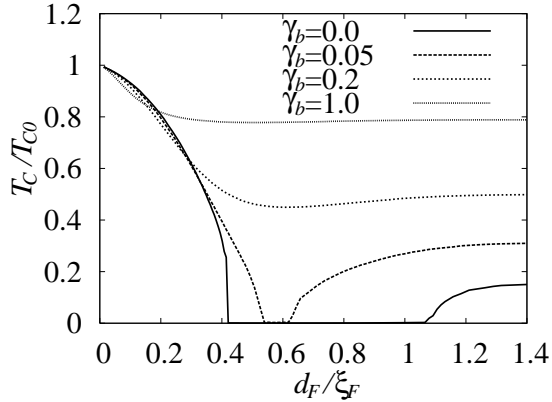} \caption{$T_C/T_{C0}$
vs.\ $d_F/\x_F$ for the various $\g_b$ without spin-flip
scattering at the interface, i.e.\ $\g_m = 0.0$.
\label{fig:tcdf1}}
\end{figure}
\begin{figure}[hbt]
\epsfxsize=8cm \epsffile{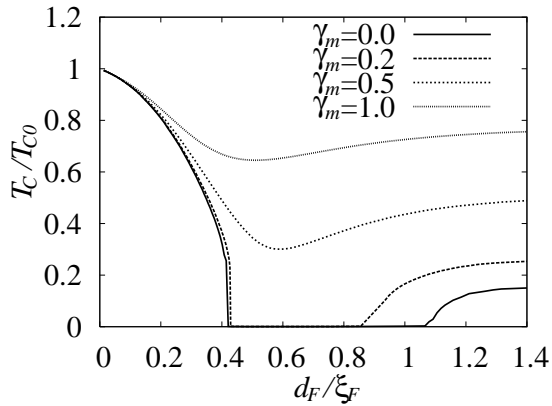} \caption{$T_C/T_{C0}$
vs.\ $d_F/\x_F$ for the various $\g_m$ with spin-flip scattering
at the interface with $\g_b = 0.0$. \label{fig:tcdf2}}
\end{figure}

\fig{fig:tcdf1} shows the effects of interface resistance on the
F thickness, $d_F$, dependence of the superconducting critical
temperature $T_C$ of S/F bilayers.
All the parameters except for $\g_b$ are same as used in \fig{fig:tcdfsf}
and the exchange field is fixed to $H_\subbox{ex} = 150$ K.
As the interface resistance parameter $\gamma_b$ is increased, the
$T_C(d_F)$ is increased and the non-monotonic feature becomes weakened. The
$T_C$ is increased because a large $\gamma_b$ decouples S from F
and, therefore, S is not influenced by F. \fig{fig:tcdf1} shows
that the non-monotonic feature is almost washed out for $\gamma_b \gtrsim
0.2$.

In \fig{fig:tcdf2}, $T_C(d_F)$ is calculated for various $\g_m$,
the interface spin flip parameter, with the other parameters same
as in \fig{fig:tcdf1}. A non-zero $\gamma_m$ induces the triplet
pairing components and couples them with the singlet component. As
can be seen from the figure, $T_C$ is enhanced by $\g_m$ as well.
Recall that $T_C$ is determined by the singlet pairing order
parameter $\Delta(x)$ in the S region, and there is no pairing
interaction in the triplet channel for the S/F system we consider
here. The triplet components come into existence only through
$\gamma_m$, and an increase of $\gamma_m$ means an increase of
the triplet pairing components. The $T_C$ is enhanced because the
singlet paring order parameter $\Delta(x)$ in the S region is
enhanced when it is coupled to the triplet components in addition
to the singlet pairing component. The exchange field of F region
does not act as a pair breaker for the triplet components unlike
for the singlet one. The detail effects of the $\gamma_m$ is
somewhat different from $\gamma_b$. When we compare two cases of
$\g_b=0.2$ in \fig{fig:tcdf1} and $\g_m=0.5$ in \fig{fig:tcdf2},
both have almost same saturated $T_C^\ast \equiv
\lim_{d_F\rarr\infty}T_C(d_F)$, but the non-monotonic structure
is much stronger in the later case than the former one. While the
$\g_m$ is similar to $\g_b$ in the sense of the increase of $T_C$
and the suppression of a minimum $T_C$ at a particular $d_F$,
they are different in that the non-monotonic structure is less
suppressed by $\g_m$ than by $\g_b$.
\begin{figure}[hbt]
\epsfxsize=8cm \epsffile{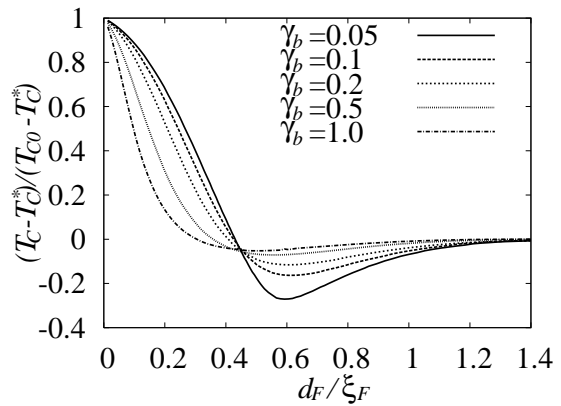}
\caption{$(T_C-T_C^\ast)/(T_{C0}-T_C^\ast)$ vs.\ $d_F/\x_F$ for
several values of the interface resistance parameter $\g_b$ with
the interface spin-flip scattering parameter $\g_m = 0.2$. As the
$\gamma_b$ is increased, the non-monotonic feature is successively
weakened. The $T_C$ increases as $\gamma_b$ is increased but
appears the other way because it is normalized as Eq.\ (\ref{Tcnormal}).
\label{fig:dipgb005}}
\end{figure}
\begin{figure}[hbt]
\epsfxsize=8cm \epsffile{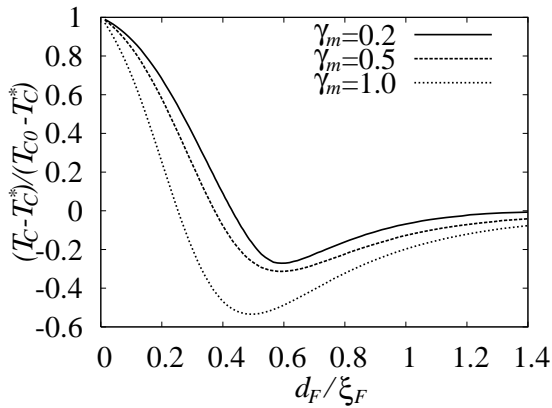}
\caption{$(T_C-T_C^\ast)/(T_{C0}-T_C^\ast)$ vs.\ $d_F/\x_F$ for
various $\g_m$ with $\g_b = 0.05$. Notice that an increase of
$\gamma_m$ enhances the non-monotonic feature.
\label{fig:dipgm02}}
\end{figure}
This can be
more clearly when the $T_C(d_F)$ is normalized in the following way:
\beq
\label{Tcnormal}
(T_C-T_C^\ast)/(T_{C0}-T_C^\ast).
 \eeq

\fig{fig:dipgb005} and \fig{fig:dipgm02} show how the
non-monotonic feature changes as $\g_b$ and $\g_m$, respectively,
are increased from $\gamma_b=0.05$ and $\gamma_m=0.2$. Both
figures were calculated with the same parameter set as in
\fig{fig:tcdf1} and \fig{fig:tcdf2} except for the $\gamma_b$ and
$\gamma_m$. The results for $\gamma_b=0.05$ and $\gamma_m=0.2$ 
are common for both figures and denoted as the solid lines. In 
\fig{fig:dipgb005}, the minimum values of the normalized $T_C$ 
increase, or the non-monotonic feature is weakened, as $\g_b$ 
increases, while the minimum values decrease due to $\g_m$ in 
\fig{fig:dipgm02}. This means that the enhancement of minimum 
$T_C$ is less strong than that of $T_C^\ast$. The reason is that 
the magnitude of the induced triplet components depend on the 
thickness of the ferromagnet of a S/F bilayer. The length scale 
over which the triplet components develop is given by $\x_F\sim 
1/k_{F}^0$ as in \eqn{eqn:kwave} which is longer than that of 
singlet component in ferromagnet given by $\x_F^{\subbox{ex}}$ of 
Eq.\ (\ref{xiF0}). Therefore, the $T_C(d_F)$ with non-zero 
$\gamma_m$, compared with $T_C(d_F)$ with $\gamma_m=0$, is 
increased more for $d_F>\x_F$ than for $d_F<\x_F$.

Another way of seeing why the magnitude of the induced triplet
components depend on the ferromagnet thickness is from the
boundary conditions of Eqs.\ (\ref{eqn:bi-bc1}) and
(\ref{eqn:bi-bc2}). The boundary condition of \eqn{eqn:bi-bc2}
means that the magnitude of the triplet components are
proportional to the derivative of the singlet component at the
interface. Since the derivative of the pairing components at
$x=-d_F$ is equal to 0 from \eqn{eqn:bi-bc1}, the derivative at
$x=0$ will be close to 0 when $d_F$ approaches 0. This
observation means that the magnitude of the triplet components
also go to 0 as $d_F$ goes to 0, and the enhancement of $T_C$ due
to the induced triplet components is not as strong for small
$d_F$ as for large $d_F$. Since the $d_F$ at which the $T_C(d_F)$
is the minimum is about $\x_F^{\subbox{ex}}$ which is smaller
than $\x_F$, the enhancement of $T_C$ for $d_F>\x_F$ gives
relative increase of the non-monotonic structure. This
enhancement of the normalized non-monotonic feature is quite
distinguishable from the effects of interface resistance and
spin-orbit scattering in bulk ferromagnet\cite{Demler97prb},
which give the suppression of the non-monotonic structure.
\section{Conclusion} \label{sec:conclusion} In this paper, we 
considered the effects of the odd-frequency $s$-wave triplet 
pairing components on the transition temperature vs.\ the 
ferromagnet thickness of S/F bilayers. The triplet pairing 
components are induced by the spin-flip scatterings at the 
interface. The interface spin-flips physically occur because the 
magnetic and non-magnetic atoms diffuse around the interface 
between the magnetic and non-magnetic layers, because the 
direction of magnetization is changed locally near the interface 
due to the boundary effects, or because the spin-orbit scattering 
at the interface induced by the polarization due to the 
difference of the work-function of two adjacent 
material.\cite{Edelstein03jetpl} The spin-flip scatterings at the 
interface were simply modeled in terms of a spin-dependent 
$\d$-function which includes potential scattering and spin-flip 
scattering. The appropriate boundary conditions describing the 
spin-flip scatterings at the interface between S and F were 
derived in the context of the Usadel equation and Eilenberger 
equation. Through the boundary condition, the singlet and triplet 
pairing components are coupled each other. The Usadel equation 
with the derived boundary conditions for S/F bilayers was then 
mapped onto an eigenvalue problem. It was solved 
self-consistently via iterations to obtain the superconducting 
transition temperature $T_C$ of S/F bilayers.

Compared with no spin-flip cases, $T_C$ is enhanced by inducing
the triplet pairing components via spin-flip scatterings at the
interface. In detail, this enhancement is distinguishable from the
enhancement of other effects, for example, the interface
resistance and the spin-orbit scattering in bulk ferromagnet.
All these
effects mostly suppress non-monotonic
dependence in $T_C$ vs.\ $d_F$, but the enhancement of $T_C$ due
to the spin flip scatterings is accompanied by
an enhancement of the relative depth of minimum $T_C$
which is scaled by the difference between
$T_{C0}$ and the saturated value, $T_C^\ast$.
This enhancement of the relative non-monotonic structure
in $T_C(d_F)$ is related to the coherence length $\x_F$ which
is the decay length of triplet pairing state in ferromagnet.
But the triplet paring components affect $T_C(d_F)$, and can be
distinguished from other effects by quantitative analysis of
$T_C(d_F)$ measurements or tunneling experiments.

We are currently extending the present analysis to the
superconductor/normal metal/ferromagnet (S/N/F) trilayer systems.
The results will be reported separately.
\begin{acknowledgments}
We would like to thank Kookrin Char and Yong-Joo Doh for having
brought this problem to our attention and for many helpful
comments and discussions. This work was supported by the Korea
Science \& Engineering Foundation (KOSEF) through grant No.\
R01-1999-000-00031-0, and by the Ministry of Education through
BK21 SNU-SKKU Physics program.
\end{acknowledgments}

\appendix
\section{Boundary conditions for Eilenberger equation}
\label{sec:BC_Eilenberger} Since the quasi-classical equations 
such as Usadel equation and Eilenberger equation describe 
variation of physical quantity over longer length scale than 
interatomic length scale, they are invaild in dealing with the 
effects of interface directly. To include the spin-flip effects 
at the interface in the quasi-classical equations, we will impose 
such effects on the boundary condition. Since the Usadel equation 
is originated from the Eilenberger equation and the Eilenberger 
equation is again from Gor'kov's equation, the derivation of the 
boundary condition consists of two steps. First, we will derive 
the boundary condition for Eilenberger equation which is the 
quasi-classical approximation of Gor'kov's equation. From 
Eilenberger equation, we will derive the boundary condition for 
Usadel equation which is the dirty limit of Eilenberger equation. 
The boundary condition for the Eilenberger equation without any 
spin-flip scattering at the interface is originally derived by 
Zaitsev\cite{Zaitsev84jetp}. Therefore, we will follow his 
procedures and then include the spin degrees of freedom and 
spin-dependent interface potential for the spin-flip scattering 
at the interface. For the insufficient part in this derivation, 
readers refer to Ref.\ \onlinecite{Zaitsev84jetp}.

First, we write down the Gor'kov's equation, which describes the 
system of two adjacent region separated by intermediate region 
with thickness $2\d$, as shown in \fig{fig:interface}. \bwide \beq 
\label{eqn:gorkov} \d(\br-\br')\d(\t-\t')\check{1} =\left( 
-\rho_z\f{\partial}{\partial \t} 
+\f{1}{2m}\f{\partial^2}{\partial \br^2} 
+\check{\bDel}-U(\br)-\check{\Si}+\m(\br) 
\right)\check{\bG}(\br,\br';\t,\t'), \eeq \ewide where 
$\check{G}$ is $4\times 4$ Green's function in Nambu-Gor'kov 
space. $\m(\br)$ is the chemical potential, and $\check{\Si}$ is 
the self-energy. Here the $U(\br)$ is the interface potential 
including spin-flip scattering, which is expressed as the 
following spin dependent delta function in spin and particle-hole 
space. \beq U(\br) = \d(z) \left( V_0\rho_0\s_0 + V_{x}\rho_0\s_x 
+ V_{y}\rho_z\s_y + V_z\rho_0\s_z \right). \eeq $\rho_i$ and 
$\s_i$ is Pauli matrix in particle-hole space, and in spin space, 
respectively. For the convenience, we  consider $V_0$ and $V_x$ 
only for potential scattering and spin-flip scattering, 
respectively, and ignore the other terms without loss of 
generality. Now we separate the above Green's function into fast 
varying part and slow varying part on $z$. \beqa 
\label{eqn:fast&slow} \check{\bG}(z,z';\rho,i\w_n) &=&
 \check{G}_{11}(z,z';\rho,i\w_n)e^{ ip_{z}(z-z')}\\
\nonumber
&&
+\check{G}_{22}(z,z';\rho,i\w_n)e^{-ip_{z}(z-z')}\\
\nonumber
&&
+\check{G}_{12}(z,z';\rho,i\w_n)e^{ ip_{z}(z+z')}\\
\nonumber && +\check{G}_{21}(z,z';\rho,i\w_n)e^{-ip_{z}(z+z')} 
\nonumber \eeqa Now $\check{G}_{ij}$ is the slowly varying 
function of $z$ and $z'$. If we substitute Eq.\ \eqn{eqn:gorkov} 
by Eq.\ \eqn{eqn:fast&slow} and neglect the second order 
derivatives, we get the equation for $\check{G}_{ij}$. \bwide 
\beq \left[-i\rho_z\w_n- (-1)^kiv_{z1}\f{\partial}{\partial z}+ 
\f{i}{2}\bv_{\parallel}\f{\partial}{\partial \brho} 
+\check{\D}(\br)-\check{\Si}\right] 
\check{G}_{kn}(z,z';\rho,i\w_n) = 0,~~~(z\neq z') \eeq \ewide 
Here, we assume the $z$-direction is normal to the interface of 
the two region and $\brho$ is the coordinates normal to the 
$z$-direction. We assume the translational symmetry in 
$\brho$-direction, ans also assume that the boundary is sharp and 
plane. When $z,z' > 0$, $v_{z1}$ is replaced by $v_{z2}$. The 
function $\check{G}_{ij}$'s for $z<z'$ and $z>z'$ are matched by 
\beq 
\check{G}_{kn}(z'+0,z')-\check{G}_{kn}(z'-0,z')=(-1)^k\f{i}{\hbar 
v_{z1(2)}} \d_{kn}. \eeq To make the equations simpler, we 
introduce $\hat{g}$ and $\hat{\cG}$ which are continuous at $z = 
z'$. \beqa \nonumber \hat{g}(z,z')   &=& 2i|v_{zj}|\left\{ 
\begin{array}{lr}
\check{G}_{11}(z,z')-\sgn(z-z'), & p_{zj} > 0,\\
\check{G}_{22}(z,z')+\sgn(z-z'), & p_{zj} < 0,\\
\end{array}
\right. \\
\hat{\cG}(z,z') &=& 2i|v_{zj}|\left\{
\begin{array}{lr}
\check{G}_{12}(z,z'), & p_{zj} > 0,\\
\check{G}_{21}(z,z'), & p_{zj} < 0.\\
\end{array}
\right.
\eeqa
For $z\ll z^\ast \equiv \mbox{min}(l_j,v_{Fj}/\bar{\ve},a)$,
we can write the equation for the Green's function using the fact
that
$\f{\partial \check{\bG}}{\partial \t}
\sim\f{\partial\check{\bG}}{\partial \t'}
\sim\bar{\ve}\check{\bG}$, where $\bar{\ve}\sim\mbox{max}(T,\D,V,\w)$ as
done by Zaitsev\cite{Zaitsev84jetp}.
The equation for Green's function $\check{\bG}$ is reduced to
\beq
H(z)\hat{\bG}(z,z') = H(z')\hat{\bG}(z,z') = \hat{0},
\eeq
where
\beq
H(z)\equiv \f{\hbar^2}{2m}\f{\partial^2}{\partial z^2}
-U(z,\brho)+\m-\f{p_{\parallel}^2}{2m}
\label{eqn:reduceH}
\eeq
for $z\neq z'$.
The solution can be written as
\beqa
\label{eqn:coef1_green}
\check{\bG} &=&
A_\pm^{(j)} e^{i p_z(z-z')} +\bar{A}_\pm^{(j)} e^{-i p_z(z-z')}\\
\nonumber
&&+B^{(j)} e^{i p_z(z+z')} +\bar{B}^{(j)} e^{i p_z(z+z')},~~~ (zz' > 0) \\
\label{eqn:coef2_green}
\check{\bG} &=&
A^{(jk)} e^{i p_z(z-z')} +\bar{A}^{(jk)} e^{-i p_z(z-z')} \\
\nonumber
&&+B^{(jk)} e^{i p_z(z+z')} +\bar{B}^{(jk)} e^{i p_z(z+z')},~~~ (zz' < 0)
\eeqa
where the superscript $j, k = 1 (2)$ denote the index of the region
$z < -\d$ ($z>\d$), and the index $+(-)$ corresponds to $z>z'$ ($z<z'$).
Now the above Green's functions can be written down
by using the two linearly independent solution of the Schr\"{o}dinger equation
$H(z)\psi_{1,2}(z) = 0$.
\beqa
\psi_1 &=& \left\{
\begin{array}{lr}
e^{ip_{z1}z}+\check{r}e^{-ip_{z1}z},& (z<0) \\
\check{d} e^{i p_{z2}z},            & (z>0)
\end{array}
\right.\\
\psi_2 &=& \left\{
\begin{array}{lr}
\f{p_{z1}}{p_{z1}}\check{d} e^{i p_{z2}z}, & (z<0) \\
e^{ip_{z1}z}-\check{r}^\dag\f{\check{d}}{\check{d}^\dag}e^{-ip_{z1}z}, & (z>0)
\end{array}
\right.
\eeqa
where $\check{r}$ and $\check{d}$ is the spin dependent coefficient.
For the spin space, it will have the form,
\beqa
\check{r} \!&\!=\!&\! \!
\bmat{cc}
\f{p_{z1}^2-V_x^2-(p_{z2}+ i V_0)^2}{(p_{z1}+p_{z2}+i V_0)^2 + V_x^2} &
\f{- 2 i p_{z1} V_x}{(p_{z1}+p_{z2}+i V_0)^2+V_x^2}
\\
\f{- 2 i p_{z1} V_x}{(p_{z1}+p_{z2}+i V_0)^2+V_x^2}&
\f{p_{z1}^2-V_x^2-(p_{z2}+ i V_0)^2}{(p_{z1}+p_{z2}+i V_0)^2 + V_x^2}
\emat,\\
\check{d} \!&\!=\!&\! \!
\bmat{cc}
\f{2p_{z1}(p_{z1}+p_{z2}+i V_0)}{(p_{z1}+p_{z2}+i V_0)^2 + V_x^2} &
\f{- 2 i p_{z1} V_x}{(p_{z1}+p_{z2}+i V_0)^2+V_x^2}
\\
\f{- 2 i p_{z1} V_x}{(p_{z1}+p_{z2}+i V_0)^2+V_x^2}&
\f{2p_{z1}(p_{z1}+p_{z2}+i V_0)}{(p_{z1}+p_{z2}+i V_0)^2 + V_x^2}
\emat.
\eeqa
For $|z|,~|z'| \ll z^\ast$, $\check{\bG}$ can be represented in the form,
\beqa
\label{eqn:trial_green}
\lefteqn{\check{\bG}(z,z')=}\\
&&\left\{
\begin{array}{lcc}
e^{ip_{z1}z'}F_{1}(z) + e^{-ip_{z1}z'}\bar{F}_{1}(z), & z'<-\d, & z>z'\\
e^{ip_{z2}z'}F_{2}(z) + e^{-ip_{z2}z'}\bar{F}_{2}(z), & z'>\d, & z<z'\\
e^{ip_{z1}z}P_{1}(z') + e^{-ip_{z1}z}\bar{P}_{1}(z'), & z<-\d, & z<z'\\
e^{ip_{z2}z}P_{2}(z') + e^{-ip_{z2}z}\bar{P}_{2}(z'), & z>\d, & z>z'
\end{array}
\right.,
\nonumber
\eeqa
where
\beqa
F_i(z) &=& \psi_1(z)\cdot\bdf_{i1} +\psi_2(z)\cdot\bdf_{i2}, \\
P_i(z) &=& \psi_1(z)\cdot\bp_{i1} +\psi_2(z)\cdot\bp_{i2},
\eeqa
Here,
$\bdf_{i1(2)}$ and $\bp_{i1(2)}$ are $4\times 4$ matrix in spin and
particle-hole space.
The coefficients $A_\pm^{j}$, $B^j$, $B^{jk}$ and $B^{jk}$ is related to
the Green's function at the vicinity of the interface. For the relation
between the Green's function of the region 1 and region 2 near the interface,
we compare the solution of Eq.\ \eqn{eqn:trial_green}
with Eqs.\ \eqn{eqn:coef1_green} and \eqn{eqn:coef2_green}, then
we can write down the coefficients $A_\pm^{j}$, $B^j$, $B^{jk}$ and $B^{jk}$
in terms of $\bdf_{i1(2)}$ and $\bp_{i1(2)}$.
If we eliminate the term $\bdf_{i1(2)}$ and $\bp_{i1(2)}$,
we get the following relations.
\beqa
\nonumber
v_1 A_{\pm}^1-v_2 A_{\pm}^2 &=&
v_1 \bar{A}_{\pm}^1-v_2 \bar{A}_{\pm}^2~= \\
v_2\f{\hat{r}\hat{d}^\dag}{\hat{d}} B^2 +v_1 \hat{r}^\dag\bar{B}^1 &=&
v_1\hat{r}B^1+v_2\f{\hat{r}^\dag\hat{d}}{\hat{d}^\dag}\bar{B}^2,
\label{eqn:rel_coef1} \\
\nonumber
v_1 \bar{A}_{\pm}^1+v_2 A_{\pm}^2 &=&
v_1 A_{\pm}^1+v_2 \bar{A}_{\pm}^2~= \\
-v_2\f{\hat{d}^\dag}{\hat{r}^\dag\hat{d}}B^2+v_1\f{1}{\hat{r}^\dag}B^1
&=&
\!\!-v_2\f{\hat{d}}{\hat{r}\hat{d}^\dag}\bar{B}^2+v_1\f{1}{\hat{r}}\bar{B}^1
\label{eqn:rel_coef2} \eeqa Here, we define the function
$\tilde{\cG}$ from $\check{\cG}$.
 \beq \tilde{\cG}=\left\{
\begin{array}{cc}
\hat{r}\check{\cG} & z < -\d,~p_i > 0 \\
\hat{r}^\dag\check{\cG} & z < -\d,~p_i < 0 \\
\f{\hat{r}\hat{d}^\dag}{\hat{d}}\check{\cG} & z >  \d,~p_i > 0 \\
\f{\hat{r}^\dag\hat{d}}{\hat{d}^\dag}\check{\cG} & z >  \d,~p_i < 0
\end{array}\right.
\eeq
Now we define the symmetric and asymmetric component of the Green's function.
\beq
\check{g}_{s(a)}=\f{\check{g}(p)\pm\check{g}(-p)}{2}
\eeq
From the Eqs. \eqn{eqn:coef1_green} and \eqn{eqn:coef2_green},
we get the following expression for $\check{g}_{s(a)}$ and $\tilde{\cG}_{s(a)}$:
\beqa
\nonumber
\check{g}_{s} &=&
                i v_j\left(A_{\pm}^j+\bar{A}_{\pm}^j\right)\\
\nonumber
\check{g}_{a} &=&
                i v_j\left(A_{\pm}^j-\bar{A}_{\pm}^j\right) - \sgn(z-z')\\
\nonumber
\tilde{\cG}_{s} &=&
                 \left\{\begin{array}{cc}
                  \f{v_1}{2}\left(
                    \hat{r}B^1+\hat{r}^\dag\bar{B}^1
                  \right) & z,z' < -\d \\
                  \f{v_2}{2}\left(
                    \f{\hat{r}\hat{d}^\dag}{\hat{d}}B^2
                   +\f{\hat{r}^\dag\hat{d}}{\hat{d}^\dag}\bar{B}^2
                  \right) & z,z' > \d
                 \end{array}\right. \\
\tilde{\cG}_{a} &=&
                 \left\{\begin{array}{cc}
                  \f{v_1}{2}\left(
                    \hat{r}B^1-\hat{r}^\dag\bar{B}^1
                  \right)& z,z' < -\d \\
                  \f{v_2}{2}\left(
                    \f{\hat{r}\hat{d}^\dag}{\hat{d}}B^2
                   -\f{\hat{r}^\dag\hat{d}}{\hat{d}^\dag}\bar{B}^2
                  \right) & z,z' > \d
                 \end{array}\right.
\label{eqn:rel_green}
\eeqa
From the relation Eqs.\ \eqn{eqn:rel_coef1}, \eqn{eqn:rel_coef2}, and
\eqn{eqn:rel_green},
we get the relations for the symmetric and the asymmetric Green's function.
\beqa
\label{eqn:bc_rel1}
\check{g}_a^{(1)} &=& \check{g}_a^{(2)} = \check{g}_a\\
\label{eqn:bc_rel2}
\tilde{\cG}_a^{(1)} &=& \tilde{\cG}_a^{(2)} = \tilde{\cG}_a\\
\label{eqn:bc_rel3}
\check{g}_s^{(1)} -\check{g}_s^{(2)}  &=&
\tilde{\cG}_s^{(1)} + \tilde{\cG}_s^{(2)}\\
\label{eqn:bc_rel4} \check{g}_s^{(1)} +\check{g}_s^{(2)}  &=&
\f{1}{R}\left(\tilde{\cG}_s^{(1)} - \tilde{\cG}_s^{(2)}\right)
\eeqa Here, $R$ is reflection coefficient. The reflection and
transmission coefficient are \beq D =
\f{p_{z2}}{p_{z1}}\hat{d}^\dag\hat{d},~~~ R = \hat{r}^\dag\hat{r}
= 1-D. \eeq For the derivation of the boundary conditions in terms
of $\check{g}$ only, we use the following relation.
 \beq
\label{eqn:conserv} \check{g}\tilde{\cG} =
(-1)^j\sgn\!\!p_j~\tilde{\cG}
 \eeq
For the proof of the above relation, the readers refer to
Zaitsev\cite{Zaitsev84jetp}. By using \eqn{eqn:conserv}, we get
the following relations. \beqa \nonumber \check{g}_s\tilde{\cG}_s
+ \check{g}_a\tilde{\cG}_a
&=& (-1)^j \tilde{\cG}_a, \\
\check{g}_s \tilde{\cG}_a + \check{g}_a\tilde{\cG}_s
&=& (-1)^j \tilde{\cG}_s.
\eeqa
Using the above equations with Eqs. \eqn{eqn:bc_rel1}, \eqn{eqn:bc_rel2},
\eqn{eqn:bc_rel3}, and \eqn{eqn:bc_rel4},
we eliminate $\tilde{\cG}_s$ and get the following relations.
\beqa
-\check{g}_s^{+}R\check{g}_s^{+}
-\left(\check{g}_s^{-}\right)^2
&=&\tilde{\cG}_a,  \\
\check{g}_s^{+}\check{g}_s^{-} +
\check{g}_s^{-}R\check{g}_s^{+}
&=&\check{g}_a\tilde{\cG}_a,   \\
\check{g}_s^+ \tilde{\cG}_a - \check{g}_a\check{g}_s^-
&=& -R\check{g}_s^+, \\
\check{g}_s^-\tilde{\cG}_a - \check{g}_aR\check{g}_s^+ &=&
-\check{g}_s^-, \eeqa where we introduce \beq \check{g}_s^\pm =
\f{\check{g}^{(1)}\pm\check{g}^{(2)}}{2}.
 \eeq
From these four relations, we again eliminate $\tilde{\cG}_a$ and
we get \beqa \check{g}_s^+ \left(\check{g}_s^{-}\right)^3 +
\check{g}_a\left(\check{g}_s^-\right)^2\!\!\!
&=&\!\!\left\{1-\left(\check{g}_s^{+}\right)^2\right\}
R\check{g}_s^+ \check{g}_s^-
\\
\left(\check{g}_s^{-}\right)^3\check{g}_s^{+}
+ \check{g}_aR\left(\check{g}_s^+\right)^2\!\!\!
&=&\!\!\check{g}_s^{-}\check{g}_s^{+}\left\{
1-R\left(\check{g}_s^{+}\right)^2\right\}
\eeqa
By considering the following commutation relations,
\beq
[\check{g}_s^+,\check{g}_s^-]_+ = \check{0},~~~\mbox{and}~~~
[\check{g}_s^\pm,\check{g}_a]_+ = \check{0},
\eeq
finally,
with Eq.\ \eqn{eqn:bc_rel1},
we get the following two boundary conditions in terms of $\check{g}$ only.
\beqa
\label{eqn:bc_eilenberger1}
\check{g}_a^{(2)}-\check{g}_a^{(1)} &=& 0, \\
\label{eqn:bc_eilenberger2}
\check{g}_a \left[
\check{R} \left(\check{g}_s^{+}\right)^2
+\left(\check{g}_s^{-}\right)^2
\right]
&=&
D\check{g}_s^{-}\check{g}_s^{+}\\
&&
-\left[R\check{g}_s^{-}\check{g}_s^{+},\left(\check{g}_s^{+}\right)^2\right]_-.
\nonumber
 \eeqa
This is the new boundary condition for the Eilenberger equation
with spin-flip scattering at the interface.
\section{Boundary conditions for Usadel equation}
\label{sec:BC_Usadel}
For the case without spin-flip scattering at the interface,
The boundary conditions for the Usadel equation is derived from
the Eilenberger equations by Kuprianov \etal\cite{Kuprianov88jetp}
in the vicinity of interface
within the mean free path $\ell$.
We will use the same methods and assumptions of Kuprianov \etal 
for deriving the boundary condition 
of Usadel equation from that of Eilenberger equation which we derived
in Appendix \ref{sec:BC_Eilenberger} for  spin-flip scattering
at the interface .
The boundary condition is directly connected to the
continuity of the normal and anomalous Green's functions of
Eilenberger equation. 
In the Eilenberger equation, if we neglect terms of $\w$ and $\D$, 
we get
\beqa
\label{eqn:simple_eilen1}
2\bel_i\cdot\nabla \check{g}_a^i 
&=& \check{g}_s^i\Av{\check{g}_s^i}-\Av{\check{g}_s^i}\check{g}_s^i \\
\label{eqn:simple_eilen2}
2\bel_i\cdot\nabla \check{g}_s^i 
&=& \check{g}_a^i\Av{\check{g}_s^i}-\Av{\check{g}_s^i}\check{g}_a^i 
\eeqa
where
$\check{g}$ is the quasi-classical Green's function in Nambu-Gor'kov space.
\beq
\check{g} = 2\pi N(0) \bmat{cc}
i \hat{g} & \hat{f} \\ 
\hat{f}^\dag & -i \hat{g}
\emat,
\eeq
with normal and anomalous Green's function in spin space,
\beq
\hat{g} = \bmat{cc} 
g_{\uparr\uparr} & g_{\uparr\dnarr} \\
g_{\dnarr\uparr} & g_{\dnarr\dnarr}
\emat,
~~\mbox{and}~~
\hat{f} = \bmat{cc} 
f_{\uparr\uparr} & f_{\uparr\dnarr} \\
f_{\dnarr\uparr} & f_{\dnarr\dnarr}
\emat
\eeq
Here, the brackets imply an integration over the solid angles,
$\Av{...} = \int d\Omega /4\pi$.
The superscript $i$ in
Eqs. \eqn{eqn:simple_eilen1} and \eqn{eqn:simple_eilen2} 
stands for the region index as
$j$ and $k$ in Eqs. \eqn{eqn:coef1_green} and \eqn{eqn:coef2_green}.
Far from the boundary, the isotropic Usadel function\cite{Usadel70prl} 
is identified as
$\check{g}_s$, and $\check{g}_a$ can be also written in terms of Usadel 
function.
\beq
\label{eqn:e_u_rel}
\check{g}_s^i = \Av{\check{g}_s^i} = \check{G}_i,~~~
\check{g}_a^i = \bel\cdot \left(\check{G}_i\nabla\check{G}_i\right),
\eeq
where
\beq
\label{eqn:def_usadel}
\check{G}=\bmat{cc}i\hat{G}&\hat{F}\\
\hat{F}^\dag&-i\hat{G}\emat.
\eeq
The derivation of the boundary condition of the Usadel equation for the 
derivatives of the Usadel function from the first boundary condition of 
Eilenberger equation, \eqn{eqn:bc_eilenberger1}, 
is the exactly same as that of Kuprianov 
\etal\cite{Kuprianov88jetp}. Therefore, we get the following boundary 
condition for the derivative of the anomalous function in Usadel equation.
\beq
\xi_1\f{\partial}{\partial z} \hat{F}_1 = \g \xi_2\f{\partial}{\partial z}
\hat{F}_2 
,~~~\g\equiv \f{\rho_1 \xi_1}{\rho_2 \xi_2}
\eeq

For the second boundary condition, we assume
$R\cdot\hat{1}\gg D(\check{g}_s^-)^2$, and $R(\check{g}_a)^2$ is 
neglected.
From Eq.\ \eqn{eqn:e_u_rel}, we have 
$\check{g}_a^i \ll \check{g}_s^i$ far enough from the boundary.
With the help of the relation $\check{g}_a^2+\check{g}_s^2=\hat{1}$, 
we simplify the Eq.\ \eqn{eqn:bc_eilenberger2}.
\beq
\label{eqn:reduce_bc_e2}
 \check{g}_a R=  \check{g}_s^-\check{g}_s^+D
\eeq From the relation between Usadel function and Eilenberger 
function, Eq.\ \eqn{eqn:e_u_rel}, we rewrite Eq.\ 
\eqn{eqn:reduce_bc_e2} in terms of Usadel function. \beq 
\ell_2\check{G}_2\nabla\check{G}_2\f{4}{3}\Av{\f{x_2R}{ D}} = 
\check{G}_2\check{G}_1 -\check{G}_1\check{G}_2 
\label{eqn:bc_usadel1} \eeq For the boundary condition of 
anomalous Usadel function, we write the Eq.\ \eqn{eqn:bc_usadel1} 
in terms of normal and anomalous Usadel function by using Eq.\ 
\eqn{eqn:def_usadel}. \beq \f{4}{3}\ell_2\hat{G}_2 \nabla 
\hat{F}\Av{\f{x_2 R}{D}} = 
[\hat{G}_2,\hat{F}_1-\hat{F}_2]_++[\hat{G}_1,\hat{F}_1-\hat{F}_2]_+ 
\label{eqn:bc_usadel2} \eeq At $T_C$, the normal Usadel function 
can be expressed as the following form which is derived in 
Appendix \ref{sec:normal_green}. \beq \hat{G}_{1(2)} = 
\left(\hat{1}+\re~\Av{\hat{r}}\right)\sgn(\w) +i\im~\Av{\hat{r}}, 
\label{eqn:normal_ftn} \eeq Now we substitute 
\eqn{eqn:normal_ftn} into \eqn{eqn:bc_usadel2}, the boundary 
condition for the anomalous Green's function will be \beqa 
\nonumber
\lefteqn{\xi_2(1+\re~\hat{r}+i\sgn(\w)\im~\hat{r})\nabla \hat{F}_2\g_B= }\\
&&\hat{F}_1-\hat{F}_2+[\re~\hat{r}+i\sgn\w\im~\hat{r},\hat{F}_1-\hat{F}_2]_+,
\eeqa where \beq \g_B = \f{2}{3}\f{\ell_2}{\xi_2}\Av{\f{x_2 
R}{D}}. \eeq Here the anomalous Usadel function is $2\times 2$ 
matrix in the spin space. This can be represented in the form of 
spin-singlet and triplet components. \beq 
\label{eqn:pauli-expand} \hat{F}(\w)= i \s_y F_{s}(+) + \s_x 
F_{tx}(-) + \s_z F_{tz}(-) + \s_0 F_{t0}(-) \eeq where \beqa 
\nonumber F_{s}  \equiv 
\f{F_{\uparr\dnarr}-F_{\dnarr\uparr}}{2},&&
F_{tx} \equiv \f{F_{\uparr\dnarr}+F_{\dnarr\uparr}}{2},\\
F_{tz} \equiv \f{F_{\uparr\uparr}-F_{\dnarr\dnarr}}{2},&&\mbox{and}~~
F_{t0} \equiv \f{F_{\uparr\uparr}+F_{\dnarr\dnarr}}{2},
\eeqa
and the plus and minus sign means even and odd function in the frequency space.
\beq
F_\a(\pm) = \f{F_\a(\w)\pm F_\a(-\w)}{2},~~~(\a=s,~tx,~tz,~\mbox{and}~t0)
\eeq
In vector notation of the anomalous Usadel function we have the following 
form of the second boundary condition.
\beq
\bcF_{1}-\bcF_{2} = \hat{\g}_b \xi_2\nabla \bcF_{2},
\eeq
where
\beqa
\bcF\equiv\bmat{c}
F_{s}(+)\\ F_{tx}(-)\\ F_{tz}(-) \\ F_{t0}(-)
\emat,~~\mbox{and}~~
\hat{\g}_b=\bmat{cccc}
\g_1 & 0 & \g_2 & 0 \\
0 & \g_3 & 0 & \g_4 \\
\g_2 & 0 & \g_1 & 0 \\
0 & \g_4 & 0 & \g_3
\emat
\label{eqn:bc_usadel_sec}
\eeqa
Since we expressed the spin-dependent interfacial potential in terms of 
$\s_0$ and $\s_x$, the following quantities are also expressed as
\beq
\av{r}=r_0\s_0+r_x\s_x,~~~\Av{\f{2}{3}\f{x_2 R}{D}}=\g_0\s_0+\g_x\s_x.
\eeq
Therefore the $\g_i$'s in \eqn{eqn:bc_usadel_sec} can be expressed as
\beqa
\g_1 = \f{1}{2}\left(\g_0 - \f{\g_x\re r_x}{1+\re r_0}\right),&&
\g_3 = \f{\g_0}{2},\\
\g_2 = \f{i}{2}\f{\g_x\im r_0-\g_0\im r_x}{1+\re r_0},&& \g_4 = 
\f{\g_x}{2}. \eeqa If we compare the result of Kuprianov when 
$r_x = 0$, we get \beq \g_0 = \f{R_b\cA}{\x_2\rho_2} \equiv \g_b 
\eeq When the interface potential $V_0$ and $V_x$ are small, and 
if we ignore the higher terms of $V_0$ and $V_x$, the parameters 
in terms of the lowest order of $V_0$ and $V_x$. \beqa \nonumber  
\g_1 \approx \g_b+\cO(V_0^2V_x^2), && \g_3 \approx \g_b,
\\
\g_2 \approx i \f{V_x}{p_{F1}+p_{F2}}\equiv i\g_m,
&&
\g_4 \approx  \cO(V_x^2) 
\eeqa
Finally, the two boundary condition for the Usadel functions are
\beqa
\xi_1\nabla \bcF_{1}=\g \xi_2\nabla \bcF_{2},&&\g = \f{\rho_1\xi_1}{\rho_2\xi_2}
\\
\bcF_{1}-\bcF_{2} = \hat{\bgam}_b \xi_2\nabla \bcF_{2},&&
\hat{\bgam_b}\approx\bmat{cccc}
 \g_b &    0 & i\g_m &   0 \\
    0 & \g_b &     0 &   0 \\
i\g_m &    0 &  \g_b &   0 \\
    0 &    0 &     0 &\g_b 
\emat.
\nonumber
\eeqa
\section{Derivation for the normal Usadel function at $T_C$}
\label{sec:normal_green} In this section, we will derive the 
Usadel function of normal Green's function at $T_C$ $\hat{G}$ of 
Eq.\ \eqn{eqn:normal_ftn} in Appendix \ref{sec:BC_Usadel}. 
Without any spin-flip scattering at the interface, $\hat{G}$ is 
simply as following. \beq \label{eqn:normal_ftn0} \hat{G} = 
\sgn\w~~~\mbox{for }V_x = 0. \eeq To get the expression for 
$\hat{G}$ with spin-flip scattering, $V_x\neq0$. We start from 
the eigen function of the Hamiltonian \eqn{eqn:reduceH}, \beqa 
\hat{\phi}_+(x)\!\!\!&=&\!\!\!\f{1}{\sqrt{2\pi }} \left\{ 
\begin{array}{ll}
\left(e^{ip_1 x}+\hat{r} e^{-ip_1 x}\right)&(x<0) \\
\hat{d} e^{ip_2 x}&(x>0)
\end{array}\right., \\
\hat{\phi}_-(x)\!\!\!&=&\!\!\!\f{\sqrt{p_1/p_2}}{\sqrt{2\pi}}\left\{
\begin{array}{ll}
\f{p_2}{p_1}\hat{d} e^{-ip_1 x}&(x<0) \\
\left(e^{-ip_2 x}-\hat{r}^\dag \f{\hat{d}}{\hat{d}^\dag} 
e^{ip_2 x}\right)&(x>0)
\end{array}\right..
\eeqa
Now,
we derive the normal Usadel function from the following definition of 
the Green's function.
\beq
\hat{G}
= \f{i}{N(0)}\sum_{p}\left\{
\begin{array}{ll} 
\f{\hat{\phi}_{+}(z_1)\hat{\phi}_{+}^\dag(z_2) 
+\hat{\phi}_{-}(z_1)\hat{\phi}_{-}^\dag(z_2)}{i\w - \e}&
(\e>0)\\
\f{\hat{\phi}_{+}^\dag(z_1)\hat{\phi}_{+}(z_2) 
+\hat{\phi}_{-}^\dag(z_1)\hat{\phi}_{-}(z_2)}{i\w - \e}&
(\e<0)
\end{array}\right.
\eeq
Here $N(0)$ is the density of state at Fermi level.
For the simplicity, we will deal with one-dimensional case.
If we write down in terms of the center of mass and relative coordinate,
we have the following form:
\bwide
\beqn
\hat{G}
&=&\f{i}{N(0)}\sum_{p}
\f{e^{i(P+p) z}+e^{-i(P+p) z}+\hat{r}^\dag e^{i(P+p)Z}+\hat{r} e^{-i(P+p) Z}}
{i\w-\e}\th(Z<z<-Z)\th(-Z)\th(\e)
\\ &&
+ \f{e^{-i(P+p) z}+e^{i(P+p) z}+\hat{r}^\dag e^{i(P+p) Z}+\hat{r} e^{-i(P+p) Z}}
{i\w-\e}\th(Z<z<-Z)\th(-Z)\th(-\e)
\\ &&
+\f{p_1}{p_2}
\f{e^{-i(P-p) z}+e^{i(P-p) z}
-\hat{r}^\dag \f{\hat{d}}{\hat{d}^\dag}e^{i(P-p) Z} 
-\hat{r}\f{\hat{d}^\dag}{\hat{d}}e^{-i(P-p) Z}}
{i\w-\e}\th(-Z<z<Z)\th(Z)\th(\e)
\\ &&
+\f{p_1}{p_2}
\f{e^{i(P-p) z}+e^{-i(P-p) z}
-\hat{r}^\dag \f{\hat{d}}{\hat{d}^\dag}e^{i(P-p) Z} 
-\hat{r}\f{\hat{d}^\dag}{\hat{d}}e^{-i(P-p) Z}}
{i\w-\e}\th(-Z<z<Z)\th(Z)\th(-\e)
\\ &&
+\f{ \hat{d}e^{-i(p Z+P z)} + \hat{d}^\dag e^{i(p Z+P z)} }
{i\w-\e}(\th(Z)\th(z<-Z)+\th(-Z)\th(z<Z))
\\&&
+\f{ \hat{d} e^{-i(p Z-P z)} + \hat{d}^\dag e^{i(p Z-P z)} }
{i\w-\e}( \th(Z)\th(z>Z) + \th(-Z)\th(z>-Z))
\eeqn
At the boundary $Z=0$, we get
\beq
\hat{G}(k_z,\w,Z=0) =\f{i}{N(0)} 
\int_{-\infty}^\infty\!dz~\sum_{P}
\re~\hat{d}\f{ e^{i P z} + e^{-iP z} } {i\w-\e}
+i\im~\hat{d} \f{ e^{iP z} - e^{-iP z} } {i\w-\e}\sgn{z}
\eeq
\ewide
After integration over $z$ and $P$, we have the normal Usadel 
function with spin-flip scattering at the interface,
\beq
\hat{G}(k_z,\w,Z=0) =
(1+\re~\av{\hat{r}})\sgn(\w)+i\im~\av{\hat{r}},
\label{eqn:normal_ftn_w_spin}
\eeq
Without any reflection at the interface,
Eq.\ \eqn{eqn:normal_ftn_w_spin} is reduced to 
to Eq.\ \eqn{eqn:normal_ftn0}.
The imaginary part in Eq.\ \eqn{eqn:normal_ftn_w_spin}
has key role to give a coupling term between
singlet pairing state of even function in frequency and
triplet pairing state of odd function in frequency.
\section{Calculation of $T_C$} \label{sec:TcCalc} In this 
section, we will get the $T_C$ equation of a S/F bilayer by 
solving Eqs.\ (\ref{eqn:bi-usadel_s}), (\ref{eqn:bi-usadel_f}), 
(\ref{eqn:bi-bc1}), (\ref{eqn:bi-bc2}), and (\ref{eqn:bi-bc3}). 
The triplet pairing components of $\bcF^{S}$ of Eq.\ 
(\ref{eqn:bi-usadel_f}) satisfy the homogeneous equation unlike 
the singlet pairing  component which has inhomogeneous term 
$\D(x)$. Therefore, the solutions of triplet components in Eq.\ 
\eqn{eqn:bi-usadel_s} with the boundary condition 
\eqn{eqn:bi-bc1} can be written down easily as follows. \beq 
\label{eqn:bi-sol_s} F_{t\a}^{S}(x)=C_{t\a}^{S}\cosh 
k_S(x-d_S),~~~(\a=x,z,\mbox{and }0) \eeq with \beq k_S = 
\f{1}{\xi_S}\sqrt{\f{|\w_n|}{\pi T_C}}. \eeq In the ferromagnetic 
region, Eq.\ \eqn{eqn:bi-usadel_f} is also homogeneous equation, 
and easily solved. The solutions of $\bcF^F$ in Eq.\ 
\eqn{eqn:bi-usadel_f} with the boundary condition 
\eqn{eqn:bi-bc1} are \beq \label{eqn:bi-sol_f} 
\bcF^{F}(x)=\bmat{cccc}
c_F(x)&c_F^\ast (x) &0 &0 \\
-c_F(x)&c_F^\ast (x) &0 &0 \\
0&0&c_{F}^0(x)&0 \\
0&0&0&c_{F}^0(x)
\emat
\bmat{c} C_{s}^{F}\\C_{tx}^{F}\\C_{tz}^{F}\\C_{t0}^{F}
\emat,
\eeq
where
\beqa
c_F(x)&=&\cosh k_F(x+d_F),\\
c_F^0(x)&=&\cosh k_F^0(x+d_F),
\eeqa
with
\beq
\label{eqn:kwave}
k_F = \f{1}{\xi_F}\sqrt{\f{|\w_n|+i h_z}{\pi T_C}},~~~\mbox{and}
k_F^0 = \f{1}{\xi_F}\sqrt{\f{|\w_n|}{\pi T_C}}.
\eeq

$F_{s}^{S}$, the singlet pairing
component of the vector $\bcF^S$, satisfies the
inhomogeneous equation \eqn{eqn:bi-usadel_s}, but the others,
the triplet pairing components, do the homogeneous equation. So, it is
useful to define projection operators to distinguish the two
group.
\beq
\label{eqn:def-project}
\cP_s
=\bmat{cccc}
1&0&0&0\\
0&0&0&0\\
0&0&0&0\\
0&0&0&0
\emat,~~~
\cP_t
=\bmat{cccc}
0&0&0&0\\
0&1&0&0\\
0&0&1&0\\
0&0&0&1
\emat .
\eeq
Here, the subscript $s$ denotes as singlet, and $t$ as triplet.
At $x=0$, from the solutions \eqn{eqn:bi-sol_s}
and \eqn{eqn:bi-sol_f}, we get the following
relations,
\beqa
\label{eqn:bc-dst-st}
\cP_t\xi_S\f{\partial}{\partial x}\bcF^{S}(0)
&=& -A_S \cP_t \bcF^{S}(0),~~~\mbox{and} \\
\label{eqn:bc-df-f}
\bcF^{F}(0)
&=&
\hat{\cB}_F\xi_F\f{\partial}{\partial x}\bcF^F(0),
\eeqa
respectively,
with
\beqa
A_S &=& k_S\xi_S\tanh k_Sd_S,~~~\mbox{and} \\
\hat{\cB}_F &=& \bmat{cccc}
\re B_F& -i\im B_F& 0 &0\\
-i\im B_F& \re B_F& 0 &0\\
0&0&B_{F0} &0\\
0&0&0&B_{F0} \emat,
\eeqa
where
\beqa
B_F &=& \left[k_F\xi_F\tanh k_Fd_F\right]^{-1}, ~\mbox{and } \\
B_{F0}&=& \left[k_F^0\xi_F\tanh k_F^0d_F\right]^{-1}. \eeqa From 
Eqs.\ \eqn{eqn:bc-df-f}, \eqn{eqn:bi-bc2}, and \eqn{eqn:bi-bc3}, 
we get the coupled boundary condition for $\bcF^S$ at $x=0$. \beq 
\xi_S\f{\partial}{\partial x}\bcF^S(0) = \hat{\cA} \bcF^S(0), 
\label{eqn:bc-ds-s} \eeq where \beq \hat{\cA} = \f{\g\hat{I}} 
{\hat{\cB}_F+\hat{\bgam}_b}. \eeq Since $\cP_s+\cP_t$ is identity 
operator, the following should be satisfied. \beq 
(\cP_s+\cP_t)\xi_S\f{\partial}{\partial x}\bcF^S(0) = 
(\cP_s+\cP_t) \hat{\cA}(\cP_s+\cP_t) \bcF^S(0). \eeq By utilizing 
\eqn{eqn:bc-dst-st} and \eqn{eqn:bc-ds-s}, we can derive the 
boundary relation between $\f{\partial}{\partial x}F^{S}_{s}(0)$ 
and $F^{S}_{s}(0)$. The boundary condition for the singlet 
component at the S region is \beq \label{eqn:bc-dss-ss} 
\xi_S\f{\partial}{\partial x}F_{s}^{S}(0) = W(\w_n)F_{s}^{S}(0), 
\eeq where \beqa \nonumber W&=&\f{ \g\left[\g+A_S(\re 
B_F+\g_b)\right]}{
A_S|B_F+\g_b|^2+\g(\re B_F+\g_b) +A_S\g_m^2\k },\\
&&\k = \f{ \g+A_S(\re B_F+\g_b)}{\g+A_S(B_{F0}+\g_b)}.
\label{eqn:ftnW-gen}
\eeqa
Here, we reduce the three component inhomogeneous difference
equation to the single component inhomogeneous differential
equation for $F^{S}_{s}(x,i\w_n)$, which satisfies
\eqn{eqn:bc-dss-ss} and
 \beq \label{eqn:bi-usadel_ss} \pi
T_C\xi_S^2\f{\partial^2}{\partial x^2} F_{s}^{S}(x,i\w_n) -|\w_n|
F_{s}^{S}(x,i\w_n) = - 2\D(x)
 \eeq
from Eq.\ \eqn{eqn:bi-usadel_s}.

To solve the inhomogeneous equation, we have to solve the
following source equation.
 \beq \label{eqn:source} \pi
T_C\xi^2\f{\partial^2}{\partial x^2} G(x,y) -|\w|G(x,y) = -
\d(x-y).
 \eeq
With the boundary condition \eqn{eqn:bc-dss-ss} and
\eqn{eqn:bi-bc1}, we have
 \beqa
\label{eqn:bc-s1}
\xi_S\f{\partial}{\partial x}G(0,y) &=&W(\w_n)G(0,y) , \\
\label{eqn:bc-s2}
\xi_S\f{\partial}{\partial x}G(d_s,y)&=&0 .
 \eeqa
The solution of Eqs.\ \eqn{eqn:source},
\eqn{eqn:bc-s1}, and \eqn{eqn:bc-s2} has the following form.
\beqa
\nonumber G(x,y,i\w_n)&=& \f{k_S/|\w_n|}
{\sinh k_Sd_s +\f{W(\w_n)}{k_S\xi_S}\cosh k_S d_s}\\
&&\times\left\{
\begin{array}{cc}
v_1(x)v_2(y),& 0<x<y \\
v_1(y)v_2(x),& y<x<d_s
\end{array}
\right. \eeqa
where
 \beqa
v_1(x)&=&\cosh k_S x +\f{W(\w_n)}{k_S\xi_S}\sinh k_S x , \\
v_2(x)&=&\cosh k_S(x-d_s).
 \eeqa
Now, the solution of Eq.\ \eqn{eqn:bi-usadel_ss} is \beq
\label{eqn:bi-ss-sol} F_{s}^{S}(x,i\w_n) = 2\int_0^{d_s}dy
G(x,y,i\w_n)\D(y).
 \eeq
Since the order parameter $\D(x)$ is determined by the symmetric
part of the anomalous function $F_s^S$, the self-consistency 
equation comes from
 \beq \label{eqn:orderp} \D(x) = \pi T_C \l g(\e_F)
\sum_{\w_n>0}F_{s}^{S}(x,i\w_n).
 \eeq
Substituting \eqn{eqn:bi-ss-sol} into \eqn{eqn:orderp} gives the
self-consistency equation,
 \beq \D(x) =2\pi T_C \l
g(\e_F)\sum_{\w_n>0}\int_0^{d_s}dy G(x,y,i\w_n)\D(y).
 \eeq
From this self-consistency equation, we can write down the
equation for $T_C$ with respect to the bulk transition
temperature $T_{C0}$.
\beqa
\label{eqn:Self_TC}
\lefteqn{\D(x)\ln\f{T_{C0}}{T_C}} \\
&=&2\pi T_C
\sum_{\w_n>0}\int_0^{d_S}\!dy
\left(\f{\d(x-y)}{|\w_n|}-G(x,y,i\w_n)\right)\D(y)
\nonumber
\eeqa
The above integral equation in \eqn{eqn:Self_TC}
can be reduced to an eigenvalue
problem after we change the integration for the function $G(x,y)$
over $y$ into summation over discrete $y$.
The integral domain $y$ is divided into $N$ grids, and we change the
integration over $y$ into the following summation to obtain Eq.\
(\ref{eqn:discrete}) in Sec.\ II.
\beq
\label{eqn:integraleqn}
\D_n\ln\f{T_{C0}}{T_{C}} =
\sum_m^N K_{nm}\D_m
\eeq
where we define
\beqa
\D_n &\equiv& \D\left(\f{d_S}{N}n\right)\\
K_{nm}&\equiv&2\pi T_C \sum_{\w_n>0}
\left( \f{\d_{nm}}{|\w_n|}-
G_{nm}(i\w_n) \right)\\
G_{nm}(i\w)&\equiv&G\left(\f{d_S}{N}n,\f{d_S}{N}m,i\w_n\right).
\eeqa
The $T_C$ is determined by the smallest eigenvalue which gives the largest
$T_C$.
\bibliographystyle{apsrev}
\bibliography{sf-bi}

\begin{thebibliography}{30}
\expandafter\ifx\csname natexlab\endcsname\relax\def\natexlab#1{#1}\fi
\expandafter\ifx\csname bibnamefont\endcsname\relax
  \def\bibnamefont#1{#1}\fi
\expandafter\ifx\csname bibfnamefont\endcsname\relax
  \def\bibfnamefont#1{#1}\fi
\expandafter\ifx\csname citenamefont\endcsname\relax
  \def\citenamefont#1{#1}\fi
\expandafter\ifx\csname url\endcsname\relax
  \def\url#1{\texttt{#1}}\fi
\expandafter\ifx\csname urlprefix\endcsname\relax\def\urlprefix{URL }\fi
\providecommand{\bibinfo}[2]{#2}
\providecommand{\eprint}[2][]{\url{#2}}

\bibitem[{\citenamefont{Werthamer}(1963)}]{Werthamer63pr}
\bibinfo{author}{\bibfnamefont{N.~R.} \bibnamefont{Werthamer}},
  \bibinfo{journal}{Phys. Rev.} \textbf{\bibinfo{volume}{132}},
  \bibinfo{pages}{2440} (\bibinfo{year}{1963}).

\bibitem[{\citenamefont{de~Gennes}(1964)}]{deGennes64rmp}
\bibinfo{author}{\bibfnamefont{P.~G.} \bibnamefont{de~Gennes}},
  \bibinfo{journal}{Rev. Mod. Phys} \textbf{\bibinfo{volume}{36}},
  \bibinfo{pages}{225} (\bibinfo{year}{1964}).

\bibitem[{\citenamefont{Fulde and Ferrel}(1964)}]{Fulde64pr}
\bibinfo{author}{\bibfnamefont{P.}~\bibnamefont{Fulde}} \bibnamefont{and}
  \bibinfo{author}{\bibfnamefont{A.}~\bibnamefont{Ferrel}},
  \bibinfo{journal}{Phys. Rev.} \textbf{\bibinfo{volume}{135}},
  \bibinfo{pages}{A550} (\bibinfo{year}{1964}).

\bibitem[{\citenamefont{Larkin and Ovchinnikov}(1965)}]{Larkin65jetp}
\bibinfo{author}{\bibfnamefont{A.}~\bibnamefont{Larkin}} \bibnamefont{and}
  \bibinfo{author}{\bibfnamefont{Y.}~\bibnamefont{Ovchinnikov}},
  \bibinfo{journal}{Sov. Phys. JETP} \textbf{\bibinfo{volume}{20}},
  \bibinfo{pages}{762} (\bibinfo{year}{1965}).

\bibitem[{\citenamefont{Bulaevskii and Kuzii}(1977)}]{Bulaevskii77sjltp}
\bibinfo{author}{\bibfnamefont{L.~N.} \bibnamefont{Bulaevskii}}
  \bibnamefont{and} \bibinfo{author}{\bibfnamefont{V.~V.} \bibnamefont{Kuzii}},
  \bibinfo{journal}{Sov. J. Low Temp. Phys.} \textbf{\bibinfo{volume}{3}},
  \bibinfo{pages}{352} (\bibinfo{year}{1977}).

\bibitem[{\citenamefont{Buzdin et~al.}(1982)\citenamefont{Buzdin, Bulaevskii,
  and Panyukov}}]{Buzdin82jetpl}
\bibinfo{author}{\bibfnamefont{A.~I.} \bibnamefont{Buzdin}},
  \bibinfo{author}{\bibfnamefont{L.~N.} \bibnamefont{Bulaevskii}},
  \bibnamefont{and} \bibinfo{author}{\bibfnamefont{S.~V.}
  \bibnamefont{Panyukov}}, \bibinfo{journal}{JETP Lett.}
  \textbf{\bibinfo{volume}{35}}, \bibinfo{pages}{178} (\bibinfo{year}{1982}).

\bibitem[{\citenamefont{Buzdin and Kupriyanov}(1990)}]{Buzdin90jetpl}
\bibinfo{author}{\bibfnamefont{A.~I.} \bibnamefont{Buzdin}} \bibnamefont{and}
  \bibinfo{author}{\bibfnamefont{M.~Y.} \bibnamefont{Kupriyanov}},
  \bibinfo{journal}{JETP Lett.} \textbf{\bibinfo{volume}{52}},
  \bibinfo{pages}{487} (\bibinfo{year}{1990}).

\bibitem[{\citenamefont{Buzdin and Kupriyanov}(1991)}]{Buzdin91jetpl}
\bibinfo{author}{\bibfnamefont{A.~I.} \bibnamefont{Buzdin}} \bibnamefont{and}
  \bibinfo{author}{\bibfnamefont{M.~Y.} \bibnamefont{Kupriyanov}},
  \bibinfo{journal}{JETP Lett.} \textbf{\bibinfo{volume}{53}},
  \bibinfo{pages}{321} (\bibinfo{year}{1991}).

\bibitem[{\citenamefont{Jiang et~al.}(1995)\citenamefont{Jiang, Davidovic,
  Reich, and Chien}}]{Jiang95prl}
\bibinfo{author}{\bibfnamefont{J.~S.} \bibnamefont{Jiang}},
  \bibinfo{author}{\bibfnamefont{D.}~\bibnamefont{Davidovic}},
  \bibinfo{author}{\bibfnamefont{D.~H.} \bibnamefont{Reich}}, \bibnamefont{and}
  \bibinfo{author}{\bibfnamefont{C.~L.} \bibnamefont{Chien}},
  \bibinfo{journal}{Phys. Rev. Lett.} \textbf{\bibinfo{volume}{74}},
  \bibinfo{pages}{314} (\bibinfo{year}{1995}).

\bibitem[{\citenamefont{M{\"{u}}hge et~al.}(1996)\citenamefont{M{\"{u}}hge,
  Garif'yanov, Goryunov, Khaliullin, Tagirov, Westerholt, Garifullin, and
  Zabel}}]{Muhge96prl}
\bibinfo{author}{\bibfnamefont{T.}~\bibnamefont{M{\"{u}}hge}},
  \bibinfo{author}{\bibfnamefont{N.~N.} \bibnamefont{Garif'yanov}},
  \bibinfo{author}{\bibfnamefont{Y.~V.} \bibnamefont{Goryunov}},
  \bibinfo{author}{\bibfnamefont{G.~G.} \bibnamefont{Khaliullin}},
  \bibinfo{author}{\bibfnamefont{L.~R.} \bibnamefont{Tagirov}},
  \bibinfo{author}{\bibfnamefont{K.}~\bibnamefont{Westerholt}},
  \bibinfo{author}{\bibfnamefont{I.~A.} \bibnamefont{Garifullin}},
  \bibnamefont{and} \bibinfo{author}{\bibfnamefont{H.}~\bibnamefont{Zabel}},
  \bibinfo{journal}{Phys. Rev. Lett.} \textbf{\bibinfo{volume}{77}},
  \bibinfo{pages}{1857} (\bibinfo{year}{1996}).

\bibitem[{\citenamefont{Kontos et~al.}(2001)\citenamefont{Kontos, Aprili,
  Lesueur, and Grison}}]{Kontos01prl}
\bibinfo{author}{\bibfnamefont{T.}~\bibnamefont{Kontos}},
  \bibinfo{author}{\bibfnamefont{M.}~\bibnamefont{Aprili}},
  \bibinfo{author}{\bibfnamefont{J.}~\bibnamefont{Lesueur}}, \bibnamefont{and}
  \bibinfo{author}{\bibfnamefont{X.}~\bibnamefont{Grison}},
  \bibinfo{journal}{Phys. Rev. Lett.} \textbf{\bibinfo{volume}{86}},
  \bibinfo{pages}{304} (\bibinfo{year}{2001}).

\bibitem[{\citenamefont{Ryazanov et~al.}(2001)\citenamefont{Ryazanov, Oboznov,
  Rusanov, Veretennikov, Golubov, and Aarts}}]{Ryazanov01prl}
\bibinfo{author}{\bibfnamefont{V.~V.} \bibnamefont{Ryazanov}},
  \bibinfo{author}{\bibfnamefont{V.~A.} \bibnamefont{Oboznov}},
  \bibinfo{author}{\bibfnamefont{A.~Y.} \bibnamefont{Rusanov}},
  \bibinfo{author}{\bibfnamefont{A.~V.} \bibnamefont{Veretennikov}},
  \bibinfo{author}{\bibfnamefont{A.~A.} \bibnamefont{Golubov}},
  \bibnamefont{and} \bibinfo{author}{\bibfnamefont{J.}~\bibnamefont{Aarts}},
  \bibinfo{journal}{Phys. Rev. Lett.} \textbf{\bibinfo{volume}{86}},
  \bibinfo{pages}{2427} (\bibinfo{year}{2001}).

\bibitem[{\citenamefont{Kontos et~al.}(2002)\citenamefont{Kontos, Aprili,
  Lesueur, Genet, Stephanidis, and Boursier}}]{Kontos02prl}
\bibinfo{author}{\bibfnamefont{T.}~\bibnamefont{Kontos}},
  \bibinfo{author}{\bibfnamefont{M.}~\bibnamefont{Aprili}},
  \bibinfo{author}{\bibfnamefont{J.}~\bibnamefont{Lesueur}},
  \bibinfo{author}{\bibfnamefont{F.}~\bibnamefont{Genet}},
  \bibinfo{author}{\bibfnamefont{B.}~\bibnamefont{Stephanidis}},
  \bibnamefont{and} \bibinfo{author}{\bibfnamefont{R.}~\bibnamefont{Boursier}},
  \bibinfo{journal}{Phys. Rev. Lett.} \textbf{\bibinfo{volume}{89}},
  \bibinfo{pages}{137007} (\bibinfo{year}{2002}).

\bibitem[{\citenamefont{Guichard et~al.}(2003)\citenamefont{Guichard, Aprili,
  Bourgeois, Kontos, Lesueur, and Gandit}}]{Guichard03prl}
\bibinfo{author}{\bibfnamefont{W.}~\bibnamefont{Guichard}},
  \bibinfo{author}{\bibfnamefont{M.}~\bibnamefont{Aprili}},
  \bibinfo{author}{\bibfnamefont{O.}~\bibnamefont{Bourgeois}},
  \bibinfo{author}{\bibfnamefont{T.}~\bibnamefont{Kontos}},
  \bibinfo{author}{\bibfnamefont{J.}~\bibnamefont{Lesueur}}, \bibnamefont{and}
  \bibinfo{author}{\bibfnamefont{P.}~\bibnamefont{Gandit}},
  \bibinfo{journal}{Phys. Rev. Lett.} \textbf{\bibinfo{volume}{90}},
  \bibinfo{pages}{167001} (\bibinfo{year}{2003}).

\bibitem[{\citenamefont{Giroud et~al.}(1998)\citenamefont{Giroud, Courtois,
  Hasselbach, Mailly, and Pannetier}}]{Giroud98prb}
\bibinfo{author}{\bibfnamefont{M.}~\bibnamefont{Giroud}},
  \bibinfo{author}{\bibfnamefont{H.}~\bibnamefont{Courtois}},
  \bibinfo{author}{\bibfnamefont{K.}~\bibnamefont{Hasselbach}},
  \bibinfo{author}{\bibfnamefont{D.}~\bibnamefont{Mailly}}, \bibnamefont{and}
  \bibinfo{author}{\bibfnamefont{B.}~\bibnamefont{Pannetier}},
  \bibinfo{journal}{Phys. Rev. B} \textbf{\bibinfo{volume}{58}},
  \bibinfo{pages}{R11872} (\bibinfo{year}{1998}).

\bibitem[{\citenamefont{Petrashov et~al.}(1999)\citenamefont{Petrashov, Sosnin,
  Cox, Parsons, and Troadec}}]{Petrashov99prl}
\bibinfo{author}{\bibfnamefont{V.~T.} \bibnamefont{Petrashov}},
  \bibinfo{author}{\bibfnamefont{I.~A.} \bibnamefont{Sosnin}},
  \bibinfo{author}{\bibfnamefont{I.}~\bibnamefont{Cox}},
  \bibinfo{author}{\bibfnamefont{A.}~\bibnamefont{Parsons}}, \bibnamefont{and}
  \bibinfo{author}{\bibfnamefont{C.}~\bibnamefont{Troadec}},
  \bibinfo{journal}{Phys. Rev. Lett.} \textbf{\bibinfo{volume}{83}},
  \bibinfo{pages}{3281} (\bibinfo{year}{1999}).

\bibitem[{\citenamefont{Gu et~al.}(2002)\citenamefont{Gu, You, Jiang, Pearson,
  Bazaliy, and Bader}}]{Gu02prl}
\bibinfo{author}{\bibfnamefont{J.~Y.} \bibnamefont{Gu}},
  \bibinfo{author}{\bibfnamefont{C.-Y.} \bibnamefont{You}},
  \bibinfo{author}{\bibfnamefont{J.~S.} \bibnamefont{Jiang}},
  \bibinfo{author}{\bibfnamefont{J.}~\bibnamefont{Pearson}},
  \bibinfo{author}{\bibfnamefont{Y.~B.} \bibnamefont{Bazaliy}},
  \bibnamefont{and} \bibinfo{author}{\bibfnamefont{S.~D.} \bibnamefont{Bader}},
  \bibinfo{journal}{Phys. Rev. Lett.} \textbf{\bibinfo{volume}{89}},
  \bibinfo{pages}{267001} (\bibinfo{year}{2002}).

\bibitem[{\citenamefont{Demler et~al.}(1997)\citenamefont{Demler, Arnold, and
  Beasley}}]{Demler97prb}
\bibinfo{author}{\bibfnamefont{E.~A.} \bibnamefont{Demler}},
  \bibinfo{author}{\bibfnamefont{G.~G.} \bibnamefont{Arnold}},
  \bibnamefont{and} \bibinfo{author}{\bibfnamefont{M.~R.}
  \bibnamefont{Beasley}}, \bibinfo{journal}{Phys. Rev. B}
  \textbf{\bibinfo{volume}{55}}, \bibinfo{pages}{15174} (\bibinfo{year}{1997}).

\bibitem[{\citenamefont{Bergeret et~al.}(2001)\citenamefont{Bergeret, Volkov,
  and Efetov}}]{Bergeret01prl}
\bibinfo{author}{\bibfnamefont{F.~S.} \bibnamefont{Bergeret}},
  \bibinfo{author}{\bibfnamefont{A.~F.} \bibnamefont{Volkov}},
  \bibnamefont{and} \bibinfo{author}{\bibfnamefont{K.~B.}
  \bibnamefont{Efetov}}, \bibinfo{journal}{Phys. Rev. Lett.}
  \textbf{\bibinfo{volume}{86}}, \bibinfo{pages}{4096} (\bibinfo{year}{2001}).

\bibitem[{\citenamefont{Edelstein}(2003)}]{Edelstein03jetpl}
\bibinfo{author}{\bibfnamefont{V.~M.} \bibnamefont{Edelstein}},
  \bibinfo{journal}{JETP Lett.} \textbf{\bibinfo{volume}{77}},
  \bibinfo{pages}{182} (\bibinfo{year}{2003}).

\bibitem[{\citenamefont{Eschrig et~al.}(2003)\citenamefont{Eschrig, Kopu,
  Cuevas, and Sch{\"{o}}n}}]{Eschrig03prl}
\bibinfo{author}{\bibfnamefont{M.}~\bibnamefont{Eschrig}},
  \bibinfo{author}{\bibfnamefont{J.}~\bibnamefont{Kopu}},
  \bibinfo{author}{\bibfnamefont{J.~C.} \bibnamefont{Cuevas}},
  \bibnamefont{and}
  \bibinfo{author}{\bibfnamefont{G.}~\bibnamefont{Sch{\"{o}}n}},
  \bibinfo{journal}{Phys. Rev. Lett.} \textbf{\bibinfo{volume}{90}},
  \bibinfo{pages}{137003} (\bibinfo{year}{2003}).

\bibitem[{\citenamefont{Volkov et~al.}(2003)\citenamefont{Volkov, Bergeret, and
  Efetov}}]{Volkov03prl}
\bibinfo{author}{\bibfnamefont{A.~F.} \bibnamefont{Volkov}},
  \bibinfo{author}{\bibfnamefont{F.~S.} \bibnamefont{Bergeret}},
  \bibnamefont{and} \bibinfo{author}{\bibfnamefont{K.~B.}
  \bibnamefont{Efetov}}, \bibinfo{journal}{Phys. Rev. Lett.}
  \textbf{\bibinfo{volume}{90}}, \bibinfo{pages}{117006}
  (\bibinfo{year}{2003}).

\bibitem[{\citenamefont{Fominov et~al.}(2003)\citenamefont{Fominov, Golubov,
  and Kupriyanov}}]{Fominov03jetpl}
\bibinfo{author}{\bibfnamefont{Y.~V.} \bibnamefont{Fominov}},
  \bibinfo{author}{\bibfnamefont{A.~A.} \bibnamefont{Golubov}},
  \bibnamefont{and} \bibinfo{author}{\bibfnamefont{M.~Y.}
  \bibnamefont{Kupriyanov}}, \bibinfo{journal}{JETP Lett.}
  \textbf{\bibinfo{volume}{77}}, \bibinfo{pages}{510} (\bibinfo{year}{2003}).

\bibitem[{\citenamefont{Bergeret et~al.}(2003)\citenamefont{Bergeret, Volkov,
  and Efetov}}]{Bergeret03prb}
\bibinfo{author}{\bibfnamefont{F.~S.} \bibnamefont{Bergeret}},
  \bibinfo{author}{\bibfnamefont{A.~F.} \bibnamefont{Volkov}},
  \bibnamefont{and} \bibinfo{author}{\bibfnamefont{K.~B.}
  \bibnamefont{Efetov}}, \bibinfo{journal}{Phys. Rev. B}
  \textbf{\bibinfo{volume}{68}}, \bibinfo{pages}{064513}
  (\bibinfo{year}{2003}).

\bibitem[{\citenamefont{Berezinskii}(1974)}]{Berezinskii74jetpl}
\bibinfo{author}{\bibfnamefont{V.~L.} \bibnamefont{Berezinskii}},
  \bibinfo{journal}{JETP Lett.} \textbf{\bibinfo{volume}{20}},
  \bibinfo{pages}{287} (\bibinfo{year}{1974}).

\bibitem[{\citenamefont{Usadel}(1970)}]{Usadel70prl}
\bibinfo{author}{\bibfnamefont{K.~D.} \bibnamefont{Usadel}},
  \bibinfo{journal}{Phys. Rev. Lett.} \textbf{\bibinfo{volume}{25}},
  \bibinfo{pages}{507} (\bibinfo{year}{1970}).

\bibitem[{\citenamefont{Fominov et~al.}(2001)\citenamefont{Fominov,
  Chtchelkatchev, and Golubov}}]{Fominov01jetpl}
\bibinfo{author}{\bibfnamefont{Y.~V.} \bibnamefont{Fominov}},
  \bibinfo{author}{\bibfnamefont{N.~M.} \bibnamefont{Chtchelkatchev}},
  \bibnamefont{and} \bibinfo{author}{\bibfnamefont{A.~A.}
  \bibnamefont{Golubov}}, \bibinfo{journal}{JETP Lett.}
  \textbf{\bibinfo{volume}{74}}, \bibinfo{pages}{96} (\bibinfo{year}{2001}).

\bibitem[{\citenamefont{Fominov et~al.}(2002)\citenamefont{Fominov,
  Chtchelkatchev, and Golubov}}]{Fominov02prb}
\bibinfo{author}{\bibfnamefont{Y.~V.} \bibnamefont{Fominov}},
  \bibinfo{author}{\bibfnamefont{N.~M.} \bibnamefont{Chtchelkatchev}},
  \bibnamefont{and} \bibinfo{author}{\bibfnamefont{A.~A.}
  \bibnamefont{Golubov}}, \bibinfo{journal}{Phys. Rev. B}
  \textbf{\bibinfo{volume}{66}}, \bibinfo{pages}{014507}
  (\bibinfo{year}{2002}).

\bibitem[{\citenamefont{Zaitsev}(1984)}]{Zaitsev84jetp}
\bibinfo{author}{\bibfnamefont{A.~V.} \bibnamefont{Zaitsev}},
  \bibinfo{journal}{Sov. Phys. JETP} \textbf{\bibinfo{volume}{59}},
  \bibinfo{pages}{1015} (\bibinfo{year}{1984}).

\bibitem[{\citenamefont{Kuprianov and Lukichev}(1988)}]{Kuprianov88jetp}
\bibinfo{author}{\bibfnamefont{M.~Y.} \bibnamefont{Kuprianov}}
  \bibnamefont{and} \bibinfo{author}{\bibfnamefont{V.~F.}
  \bibnamefont{Lukichev}}, \bibinfo{journal}{Sov. Phys. JETP}
  \textbf{\bibinfo{volume}{67}}, \bibinfo{pages}{1163} (\bibinfo{year}{1988}).

\end{thebibliography}
\end{document}